# First-Principles Investigation of the Structural, Electronic, Optical, Elastic, and Thermoelectric Properties of $X_2MoSeO_6$ (X = K, Rb, Cs)

Sha Ajharul Islam Forhat, Md. Rakibul Hasan Rakib, Md. Al Noman Ruhin, and Md Tashfiq Bin Kashem*

Department of Electrical and Electronic Engineering, Ahsanullah University of Science and Technology, Dhaka - 1208, Bangladesh

* Corresponding author: E-mail address: tashfiq.eee@aust.edu

**Abstract**

Understanding how alkali-metal substitution influences the coupled physical properties of molybdenum selenite oxides is essential for establishing their composition-structure-property relationships. Although the lighter members of the $X_2MoSeO_6$ family have been investigated, the heavier alkali-metal compounds remain comparatively unexplored, particularly regarding their integrated electronic, mechanical, optical, and thermoelectric responses. Here, a comprehensive first-principles investigation of $X_2MoSeO_6$ (X = K, Rb, and Cs) is presented using density functional theory and semi-classical Boltzmann transport calculations. The optimized structures exhibit progressive unit-cell expansion with increasing alkali-metal ionic size, while their indirect electronic bandgaps vary non-monotonically, reaching 2.646, 2.216, and 2.427 eV for $K_2MoSeO_6$, $Rb_2MoSeO_6$, and $Cs_2MoSeO_6$, respectively. All three structures satisfy the elastic stability criteria, with Hill-averaged bulk moduli increasing from 16.332 to 20.975 GPa across the series. $K_2MoSeO_6$ exhibits the strongest elastic anisotropy, whereas $Rb_2MoSeO_6$ possesses the highest static electronic dielectric constant (4.614) and maximum absorption coefficient ($2.291 \times 10^6$ cm$^{-1}$). Thermoelectric calculations reveal pronounced carrier-polarity-dependent behavior, with $Rb_2MoSeO_6$ attaining a positive Seebeck coefficient of 390.55 μV K$^{-1}$ at 100 K and an electronic-only figure of merit ($ZT_e$) of 7.286 at 300 K under the selected positive-Seebeck condition. The contrasting trends in bandgap, mechanical anisotropy, optical absorption, and electronic transport demonstrate that alkali-metal substitution affects these properties through distinct structural and electronic mechanisms rather than a uniform compositional dependence. By extending the investigation to the K-Rb-Cs series and integrating multiple physical responses, this work establishes a comparative foundation for understanding and further exploring the functional characteristics of molybdenum selenite oxides.

## 1. Introduction

The relationship between crystal chemistry and functional properties is central to the design of multicomponent inorganic materials. In oxide compounds containing transition-metal and main-group cations, variations in ionic size, coordination geometry, and polyhedral connectivity can substantially alter electronic structure, mechanical response, and optical behavior [1]–[3]. Alkali-metal molybdenum selenites provide an interesting system in this regard. Their structures contain Mo(VI)-O and Se(IV)-O coordination units, while the alkali-metal constituents influence the arrangement and connectivity of these units. Such compositional sensitivity offers an opportunity to investigate how structural changes are reflected in the physical properties of chemically related compounds.

Eagleman and Halasyamani [4] experimentally synthesized and characterized $A_2SeMoO_6$ (A = Na, K, and Rb), revealing pronounced structural differences despite the compounds having the same stoichiometric form. $Na_2SeMoO_6$ exhibits a three-dimensional cubic structure, whereas $K_2SeMoO_6$ crystallizes in the monoclinic $P2_1/c$ space group with molecular structural units and $Rb_2SeMoO_6$ adopts an orthorhombic *Pnma* structure with one-dimensional connectivity. These findings established the importance of alkali-metal substitution in controlling the structural organization of Mo(VI)-Se(IV) oxides. However, the experimental investigation did not systematically address the consequences of these structural differences for electronic, elastic, optical, or thermoelectric properties.

Recent first-principles investigations have demonstrated the potential of composition-dependent studies in related molybdenum-containing and selenium-based compounds. Khan *et al.* [5] examined $Na_2MoXO_6$ (X = Si, Ge, and Sn), reporting semiconducting electronic structures and substantial variations in optical and mechanical properties with chemical substitution. Li *et al.* [6] investigated $A_2SeX_6$ (A = K, Rb; X = Cl, Br, and I), showing that changes in alkali-metal and halogen composition influence bandgap energies, mechanical characteristics, and optical absorption. Related calculations on $K_2SeX_6$ (X = Cl and Br) [7] and $Rb_2SeX_6$ (X = Cl, Br, and I) [8] further demonstrated the sensitivity of electronic and optical properties to compositional changes. Zahan *et al.* [9] also examined the pressure-dependent structural, mechanical, optoelectronic, and thermoelectric responses of $K_2SeCl_6$. Although these selenium-containing halides differ from molybdenum selenite oxides in bonding and structural topology, they provide useful examples of how chemical composition and structural modification affect coupled physical properties.

A more direct investigation was reported by Abdalla *et al.* [10], who employed density functional theory to examine the structural, mechanical, electronic, and optical properties of $X_2MoSeO_6$ (X = Li and Na). Their calculations identified semiconducting behavior and composition-dependent optical characteristics, extending the available knowledge of this compound family beyond experimental crystallography. Nevertheless, the heavier alkali-metal members remain comparatively less explored from a computational perspective. In particular, a systematic comparison of $K_2MoSeO_6$, $Rb_2MoSeO_6$, and $Cs_2MoSeO_6$ is needed to establish how their different structural environments influence electronic states, elastic anisotropy, optical excitations, and thermoelectric transport.

These properties are closely interconnected. Changes in lattice geometry can modify orbital interactions and electronic-band dispersion, thereby influencing the energies and strengths of optical transitions [3], [11], [12]. Similarly, the elastic stiffness tensor provides information on resistance to deformation, mechanical stability, and directional anisotropy [13]–[16]. Thermoelectric transport introduces another important dimension because the Seebeck coefficient, electrical conductivity, and electronic thermal conductivity depend on the distribution of electronic states and their contributions to charge and heat transport [17]–[20]. Examining these responses together therefore offers a more informative description of composition-dependent behavior than treating each property independently.

The present study investigates the structural, electronic, elastic, optical, and thermoelectric properties of $X_2MoSeO_6$ (X = K, Rb, and Cs) using first-principles calculations and semiclassical Boltzmann transport theory [20]–[22]. The experimentally reported monoclinic and orthorhombic structures are considered for $K_2MoSeO_6$ and $Rb_2MoSeO_6$, respectively, while $Cs_2MoSeO_6$ is investigated in an orthorhombic *Pnma* configuration. The analysis encompasses optimized crystallographic parameters, electronic band structures and orbital-resolved densities of states, elastic stiffness and anisotropy, frequency-dependent optical response, and temperature-dependent thermoelectric coefficients under contrasting carrier-polarity conditions.

The principal contribution of this work is a unified comparison of the K-Rb-Cs series that extends the existing experimental and computational literature on alkali-metal molybdenum selenites. Particular attention is given to the non-monotonic variation in electronic bandgap, the evolution of elastic anisotropy, the redistribution of high-energy optical response, and the dependence of thermoelectric behavior on carrier occupation. By relating these properties to alkali-metal substitution and structural symmetry, this study provides further insight into the composition-structure-property relationships of Mo-Se-O compounds.

## 2. Computational Methodology

First-principles density functional theory (DFT) calculations were performed to investigate the structural, electronic, elastic, and optical properties of $X_2MoSeO_6$ (X = K, Rb, and Cs) using the Cambridge Serial Total Energy Package (CASTEP) within the Materials Studio environment [21]. Exchange-correlation effects were described using the Perdew-Burke-Ernzerhof functional revised for solids (PBEsol) within the generalized gradient approximation (GGA) [2]. The optimized crystal geometries provided a common structural basis for analyzing the composition-dependent mechanical and optoelectronic responses. Thermoelectric transport was investigated separately using electronic structures obtained with Quantum ESPRESSO and processed with BoltzTraP2 [20], [22], [23].

All three crystal structures were fully relaxed with respect to lattice parameters and internal atomic coordinates using the Broyden-Fletcher-Goldfarb-Shanno (BFGS) quasi-Newton minimization scheme. The geometry-optimization convergence tolerances were $2 \times 10^{-5}$ eV for the total-energy change, 0.05 eV $Å^{-1}$ for the maximum residual force, 0.1 GPa for the maximum residual stress, and 0.002 Å for the maximum atomic displacement. Full unit-cell optimization was carried out with fixed basis quality and a maximum of 100 geometry-optimization steps. These criteria were applied to obtain equilibrium configurations for subsequent property calculations.

In CASTEP, the electronic wave functions were expanded in a plane-wave basis with a kinetic-energy cutoff of 600 eV. Brillouin-zone sampling employed a 4 × 4 × 4 Monkhorst-Pack k-point mesh [24]. The self-consistent field (SCF) energy convergence tolerance was $2 \times 10^{-6}$ eV $atom^{-1}$, with up to 100 SCF iterations and a convergence window of three successive iterations. Electronic minimization used the All Bands/EDFT scheme, while electron-ion interactions were represented by norm-conserving pseudopotentials with scalar-relativistic effects treated using the Koelling-Harmon approach [25].

Electronic properties were characterized from the calculated band structures, total density of states (TDOS), and orbital-projected density of states (PDOS), allowing the band-edge energies and orbital contributions to be compared across the series. The elastic stiffness tensors $C_{ij}$ were used to assess mechanical stability and to determine effective bulk, shear, and Young's moduli and Poisson's ratios using Voigt-Reuss-Hill averaging [13], [14]. Directional variations in Young's modulus, linear compressibility, shear modulus, and Poisson's ratio were visualized using ELATE [16]. Optical response functions were derived from the frequency-dependent complex dielectric function, including the refractive index, extinction coefficient, absorption coefficient, reflectivity, optical conductivity, and electron energy-loss function.

For thermoelectric analysis, electronic band-structure data were generated using the Quantum ESPRESSO DFT package [22], [23], and the semiclassical Boltzmann transport coefficients were evaluated with BoltzTraP2 [20] within the constant relaxation-time approximation (CRTA). The Seebeck coefficient S, electrical conductivity divided by relaxation time $\sigma/\tau$, electronic thermal conductivity divided by relaxation time $\kappa_e/\tau$, and power factor divided by relaxation time $PF/\tau$ were calculated over 100-800 K. The electronic-only thermoelectric figure of merit was evaluated as $ZT_e = S^2\sigma T/\kappa_e$, distinguishing the electronic contribution from the full figure of merit that also includes lattice thermal conductivity.

### 3.1. Structural Properties and Stability

The equilibrium crystal structures of $X_2MoSeO_6$ (X = K, Rb, and Cs) were optimized using density functional theory (DFT) within the PBEsol-GGA approximation. The optimized lattice parameters, unit-cell volumes, and densities are summarized in Tables 1 and 2, together with available experimental crystallographic data. The optimized crystal structures were visualized using VESTA (Fig. 1). The structural analysis provides insight into how alkali-metal substitution influences lattice geometry, crystallographic symmetry, and the spatial arrangement of the Mo-Se-O framework, establishing a basis for understanding the composition-dependent physical properties of these compounds.

$K_2MoSeO_6$ crystallizes in a monoclinic structure with the $P2_1/c$ space group (No. 14), whereas $Rb_2MoSeO_6$ adopts an orthorhombic structure belonging to the *Pnma* space group (No. 62), consistent with the experimental findings of Porter and Halasyamani [4]. The $Cs_2MoSeO_6$ structure investigated in the present work was modeled in the orthorhombic Pnma configuration. In the monoclinic $K_2MoSeO_6$ structure, the constituent atoms occupy the 4e Wyckoff positions, while the orthorhombic structures contain atoms distributed over the 4c and 8d sites.

The change in crystal symmetry from monoclinic $K_2MoSeO_6$ to orthorhombic $Rb_2MoSeO_6$ demonstrates that alkali-metal substitution affects not only lattice dimensions but also the geometric arrangement and symmetry of the constituent polyhedral units. Such structural changes can modify interatomic distances, bond angles, and orbital interactions, thereby influencing the elastic and electronic responses of the materials.

Table 1. Optimized crystallographic parameters of $X_2MoSeO_6$ (X = K, Rb, and Cs) compared with available experimental data.

| Compound | Source | Crystal system / space group | a (Å) | b (Å) | c (Å) | β (°) |
|---|---|---|---|---|---|---|
| $K_2MoSeO_6$ | Present work | Monoclinic, $P2_1/c$ | 6.203 | 15.375 | 7.723 | 115.79 |
| | Experiment [4] | Monoclinic, $P2_1/c$ | 6.118 | 15.395 | 7.580 | 112.39 |
| $Rb_2MoSeO_6$ | Present work | Orthorhombic, *Pnma* | 7.924 | 6.215 | 14.555 | 90 |
| | Experiment [4] | Orthorhombic, *Pnma* | 7.805 | 6.188 | 14.405 | 90 |
| $Cs_2MoSeO_6$ | Present work | Orthorhombic, *Pnma* | 8.076 | 6.526 | 15.009 | 90 |

The optimized lattice parameters exhibit good agreement with the available experimental results. For $K_2MoSeO_6$, the calculated lattice constants deviate from the experimental values by +1.39%, −0.13%, and +1.89% along the a, b, and c directions, respectively. The optimized monoclinic angle of 115.79° is 3.40° larger than the experimental value of 112.39°. Similarly, the calculated lattice parameters of $Rb_2MoSeO_6$ differ from the experimental values by +1.52%, +0.44%, and +1.04%, respectively.

The relatively small deviations in lattice constants demonstrate that the PBEsol-GGA calculations reproduce the principal crystallographic dimensions of the experimentally characterized compounds with good accuracy. This agreement supports the use of the optimized structures as a basis for investigating their elastic, electronic, and optical properties.

The equilibrium unit-cell volume increases systematically with alkali-metal ionic size, from 663.162 Å$^3$ for $K_2MoSeO_6$ to 716.799 Å$^3$ for $Rb_2MoSeO_6$ and 791.034 Å$^3$ for $Cs_2MoSeO_6$. The calculated volumes of $K_2MoSeO_6$ and $Rb_2MoSeO_6$ agree well with the corresponding experimental values of 660.26 and 695.90 Å$^3$, respectively [4], with minor deviations of approximately 0.44% and 3.00%.

Relative to $K_2MoSeO_6$, the unit-cell volumes of $Rb_2MoSeO_6$ and $Cs_2MoSeO_6$ are larger by approximately 8.09% and 19.28%, respectively. This progressive expansion reflects the greater spatial requirements of the larger alkali-metal cations and the associated adjustments in local coordination geometry. The resulting structural differences may also contribute to the composition-dependent electronic and mechanical responses discussed in subsequent sections.

Table 2. Calculated unit-cell volumes, densities, and geometric descriptors of $X_2MoSeO_6$.

| Parameter | $K_2MoSeO_6$ | $Rb_2MoSeO_6$ | $Cs_2MoSeO_6$ |
|---|---|---|---|
| Unit-cell volume, $V_o$ (Å$^3$) | 663.162 | 716.799 | 791.034 |
| Reported density (g cm$^{-3}$) | 3.49 | 4.09 | 4.49 |
| Tolerance factor, $\tau_G$ | 1.01 | 1.06 | 1.12 |
| Octahedral factor, $\mu$ | 0.41 | 0.41 | 0.41 |
| $\Delta\mu$ | 0.03 | 0.03 | 0.03 |

The reported density increases from 3.49 to 4.49 g cm$^{-3}$ across the K-Rb-Cs series despite the simultaneous expansion of the unit cell. For a crystalline solid, density depends on the mass and volume of the unit cell according to $\rho = ZM/(N_aV_0)$, where Z is the number of formula units per unit cell, M is the molar mass, $N_a$ is Avogadro's constant, and $V_0$ is the unit-cell volume.

Although larger alkali-metal ions promote lattice expansion, their greater atomic masses increase the mass contained within the unit cell. The competition between mass increase and volume expansion governs the overall density trend.

The influence of alkali-metal substitution on crystal geometry is further reflected in the supplied tolerance and octahedral factors. The tolerance factor increases from 1.01 for $K_2MoSeO_6$ to 1.06 for $Rb_2MoSeO_6$ and 1.12 for $Cs_2MoSeO_6$, whereas the octahedral factor remains unchanged at 0.41.

The increasing tolerance factor reflects changes in the relative ionic sizes and geometric accommodation associated with larger alkali-metal cations. The unchanged octahedral factor follows from the unchanged nominal ions defining the Mo-centered coordination descriptor. As these structures are not all conventional corner-sharing $ABX_3$ perovskites, these geometric ratios are best interpreted as comparative size descriptors rather than direct measures of thermodynamic stability.

Taken together, the results reveal a clear composition-dependent structural evolution across $X_2MoSeO_6$. Substitution of K by Rb and Cs produces systematic unit-cell expansion and changes in the reported density, while the experimentally established monoclinic-to-orthorhombic change highlights the sensitivity of the framework to alkali-metal chemistry. The agreement between calculated and experimental lattice parameters provides a sound structural foundation for interpreting the electronic, optical, elastic, and thermoelectric trends of the investigated compounds.

### 3.2. Electronic Properties

The electronic properties of $K_2MoSeO_6$, $Rb_2MoSeO_6$, and $Cs_2MoSeO_6$ were investigated through their PBEsol-GGA electronic band structures, TDOS, and PDOS, as shown in the corresponding band-structure and density-of-states panels (Figs. 2-4). Together, these results elucidate the evolution of the electronic bandgap, the distribution of density of states, and the orbital contributions governing the electronic structures of the investigated compounds.

The calculated band structures reveal indirect semiconducting gaps of 2.646, 2.216, and 2.427 eV for $K_2MoSeO_6$, $Rb_2MoSeO_6$, and $Cs_2MoSeO_6$, respectively. Substituting K with Rb reduces the bandgap by 0.430 eV, whereas further substitution with Cs increases it by 0.211 eV, making $Rb_2MoSeO_6$ the narrowest-gap compound. Interestingly, this non-monotonic behavior contrasts with the progressive expansion of the optimized unit-cell volume from 663.162 Å$^3$ (K) to 716.799 Å$^3$ (Rb) and 791.034 Å$^3$ (Cs). This non-monotonic evolution reveals that the bandgap is governed by the combined effects of lattice geometry and the relative energies and dispersions of framework-derived electronic states, rather than by unit-cell expansion alone.

The band extrema identified along the calculated high-symmetry paths place the conduction-band minimum at Γ for all three compounds, whereas the valence-band maxima lie along the M-K, S-X, and Y-S segments for $K_2MoSeO_6$, $Rb_2MoSeO_6$, and $Cs_2MoSeO_6$, respectively. This separation of the band-edge wavevectors establishes the indirect-gap character of the plotted electronic dispersions.

The band dispersions exhibit several relatively flat electronic branches within both the valence and conduction manifolds, particularly near the band edges. Such weakly dispersive states contribute to pronounced DOS features because electronic states are concentrated within relatively narrow energy intervals. This correspondence is evident in the TDOS spectra, which display multiple peaks associated with groups of electronic bands. From a transport perspective, reduced band curvature may indicate relatively large carrier effective masses along particular crystallographic directions, potentially affecting carrier mobility. The relation $m^{*-1} = \hbar^{-2}\partial^2E/\partial k^2$ links the band-edge curvature to the carrier effective mass along a chosen reciprocal-space direction; hence, the observed differences in dispersion suggest a possible route to composition-dependent transport behavior. The differences in TDOS peak positions and intensities among the compounds further indicate that alkali-metal substitution modifies the electronic-state distribution rather than merely shifting an otherwise unchanged spectrum.

The PDOS profiles reveal a broadly similar orbital organization across the three compositions. O-p states dominate the upper valence manifold, producing prominent features between approximately −2 and 0 eV in the supplied DOS energy coordinates. The strongest O-p features occur near −0.7 eV, illustrating the concentration of oxygen-derived states close to the valence-band edge. Deeper valence states, particularly between −6 and −2 eV, contain appreciable contributions from Mo-d and Se-p orbitals alongside O-p states. A notable Mo-d feature appears near −3 eV in all three compounds, overlapping with substantial oxygen-derived spectral weight. The coincident O-p, Se-p, and Mo-d spectral features are consistent with orbital mixing within the Mo-Se-O framework, linking the electronic structure to the coordination environment established by structural optimization.

The lower conduction manifold exhibits a distinctly different orbital character, with Mo-d states providing the dominant contribution to the first pronounced conduction-region peak, accompanied by appreciable O-p spectral weight. This feature occurs near 2.8-3.1 eV in the supplied energy coordinates and is especially prominent in $K_2MoSeO_6$, where the Mo-d contribution is especially pronounced compared with the corresponding $Rb_2MoSeO_6$ and $Cs_2MoSeO_6$ features. The contrast between the oxygen-rich valence states and molybdenum-rich conduction states suggests that low-energy electronic excitations involve substantial changes in orbital character within the Mo-Se-O framework. The joint O-p and Mo-d character is particularly relevant to the interband transition pathways contributing to the calculated optical response, although optical intensities additionally depend on transition matrix elements [3].

The alkali-metal-derived orbitals contribute relatively little to the principal valence and low-energy conduction features, although distinct contributions appear at deeper occupied energies and higher conduction energies. In particular, Rb-p and Cs-p states produce pronounced features in the deeper valence region, while Cs-d contributions become noticeable toward the upper end of the plotted conduction-energy range. These observations suggest that the alkali-metal constituents influence the electronic structure primarily through their effects on the framework environment rather than by introducing dominant states at the fundamental band edges.

The indirect nature of the fundamental bandgaps is particularly relevant to the optical response. Because the valence-band maximum and conduction-band minimum occur at different wavevectors, transitions between these extrema generally require phonon assistance to satisfy crystal-momentum conservation [26], [27]. Consequently, weak phonon-assisted absorption may occur near the indirect bandgap, whereas stronger optical absorption is governed primarily by the energies and transition matrix elements of dipole-allowed direct interband transitions [26]. Thus, the fundamental electronic bandgap does not necessarily coincide with the onset of strong optical absorption. These electronic-structure characteristics provide a basis for interpreting the dielectric response and absorption spectra discussed in the following section.

Overall, alkali-metal substitution provides a clear means of tuning the electronic structure, yielding a non-monotonic bandgap sequence with a minimum of 2.216 eV for $Rb_2MoSeO_6$. Despite their structural differences, all three compounds share a characteristic electronic organization dominated by O-p states in the upper valence manifold and Mo-d states in the lower conduction manifold. The persistence of these orbital features alongside the variation in bandgap energy highlights the sensitivity of the Mo-Se-O electronic framework to alkali-metal substitution and provides a foundation for understanding the composition-dependent optical properties.

### 3.3. Optical Properties

The optical properties of $X_2MoSeO_6$ (X = K, Rb, Cs) were investigated to examine how alkali-metal substitution modifies electronic polarization, interband excitation, and photon attenuation. Figure 5 (a-h) presents the real and imaginary dielectric functions, refractive index, absorption coefficient, reflectivity, real and imaginary optical conductivities, and electron energy-loss function over the photon-energy range of 0-30 eV. The analysis is based on the frequency-dependent electronic dielectric response, from which the remaining optical parameters are obtained.

#### 3.3.1. Dielectric Function

The complex dielectric function describes the response of a material to an oscillating electric field and is expressed as [3], [12], [28]

$$\epsilon(\omega) = \epsilon_1(\omega) + j\epsilon_2(\omega) \quad (1)$$

where $\epsilon_1$ describes the dispersive electronic polarization and $\epsilon_2$ represents the absorptive response associated with electronic excitations. As shown in Fig. 5(a), the static electronic dielectric constants, $\epsilon_1(0)$, are 4.038, 4.614, and 4.510 for $K_2MoSeO_6$, $Rb_2MoSeO_6$, and $Cs_2MoSeO_6$, respectively. $Rb_2MoSeO_6$ therefore exhibits the strongest low-frequency electronic screening, followed closely by $Cs_2MoSeO_6$. This ordering is broadly consistent with the calculated indirect bandgaps of 2.646, 2.216, and 2.427 eV, respectively. Nevertheless, the dielectric constant is determined by the combined contributions of optically allowed transitions throughout the electronic structure and cannot be inferred from the fundamental bandgap alone. The close correspondence between the dielectric screening order and the bandgap trend is consistent with the greater polarizability generally associated with lower-energy interband excitations; the full spectrum of transition energies and oscillator strengths determines the precise magnitude [11], [28].

The real dielectric function reaches its principal maximum at 2.728 eV for $K_2MoSeO_6$, 2.745 eV for $Rb_2MoSeO_6$, and 2.846 eV for $Cs_2MoSeO_6$, with corresponding magnitudes of 5.555, 6.359, and 6.298. These closely spaced maxima suggest similar low-energy polarization mechanisms, while the differences in intensity indicate changes in electronic transition strengths following alkali-metal substitution. $K_2MoSeO_6$ retains a positive real dielectric response throughout the investigated energy range. In contrast, $\epsilon_1$ becomes negative over limited intervals for $Rb_2MoSeO_6$ and $Cs_2MoSeO_6$ at higher photon energies, reaching minima of −0.746 at 19.92 eV and −0.833 at 16.88 eV, respectively. A negative real permittivity signifies that the electronic polarization response is out of phase with the applied field sufficiently to yield $\epsilon_1 < 0$. In these spectral windows, electromagnetic fields may become strongly attenuated and interface reflection can be enhanced, with the response also controlled by $\epsilon_2$. The occurrence of negative $\epsilon_1$ at high photon energies is therefore a signature of a metal-like optical response within a restricted frequency range, rather than metallic dc transport [28], [29].

The imaginary dielectric spectra in Fig. 5(b) provide a complementary description of interband excitations. $K_2MoSeO_6$ exhibits its strongest peak at 6.702 eV, with $\epsilon_2$=3.956, together with a lower-energy feature near 3.54 eV. $Rb_2MoSeO_6$ and $Cs_2MoSeO_6$ show their principal maxima at 3.720 and 3.712 eV, reaching 4.030 and 3.834, respectively. Additional features appear at higher energies, particularly near 18.9 eV for $Rb_2MoSeO_6$ and 13.8 eV for $Cs_2MoSeO_6$. The electronic structure provides a basis for interpreting these transitions. The upper valence bands are dominated by O-p states, whereas Mo-d orbitals contribute substantially to the lower conduction bands. Transitions involving these states are therefore expected to influence the initial optical response, with transition intensities governed by the optical dipole matrix elements.

### 3.3.2. Refractive Index and Extinction Coefficient

The refractive index and extinction coefficient are obtained from the complex dielectric function through the complex refractive index,

$$N(\omega) = n(\omega) + jk(\omega) = \sqrt{\epsilon(\omega)} \tag{2}$$

where n determines the phase propagation of electromagnetic waves and k characterizes their attenuation. For a nonmagnetic material, the two quantities are calculated as [28], [29]

$$n(\omega) = \sqrt{\frac{\sqrt{\epsilon_1^2(\omega) + \epsilon_2^2(\omega)} + \epsilon_1(\omega)}{2}} \tag{3}$$

$$k(\omega) = \sqrt{\frac{\sqrt{\epsilon_1^2(\omega) + \epsilon_2^2(\omega)} - \epsilon_1(\omega)}{2}} \tag{4}$$

In the nonabsorbing static limit, these relations reduce to

$$n(0) = \sqrt{\epsilon_1(0)} \tag{5}$$

The calculated static refractive indices are 2.009, 2.148, and 2.124 for $K_2MoSeO_6$, $Rb_2MoSeO_6$, and $Cs_2MoSeO_6$, respectively, consistent with their static electronic dielectric constants.

As illustrated in Fig. 5(c), the refractive indices initially increase with photon energy, reaching maxima of 2.378 at 2.807 eV, 2.553 at 2.864 eV, and 2.539 at 2.955 eV for the K-, Rb-, and Cs-based compounds, respectively. The stronger dispersion of $Rb_2MoSeO_6$ and $Cs_2MoSeO_6$ follows their enhanced low-energy dielectric response. Beyond the principal maxima, the refractive indices decrease and exhibit additional features associated with higher-energy electronic transitions.

At elevated photon energies, the refractive indices fall below unity, reaching minima of approximately 0.605, 0.393, and 0.375 for $K_2MoSeO_6$, $Rb_2MoSeO_6$, and $Cs_2MoSeO_6$, respectively. These values reflect anomalous high-energy dispersion; a phase refractive index below unity does not imply superluminal signal or energy transmission [28], [29].

### 3.3.3. Absorption Coefficient

The absorption coefficient describes the exponential attenuation of light propagating through a material. It is related to the extinction coefficient by [28], [29]

$$\alpha(\omega) = \frac{2\omega k(\omega)}{c} = \frac{4\pi k(\omega)}{\lambda} \tag{6}$$

where c is the speed of light in vacuum and λ is the corresponding vacuum wavelength. Combining Eqs. (4) and (6) gives

$$\alpha(\omega) = \frac{\omega\sqrt{2}}{c}\sqrt{\sqrt{\epsilon_1^2(\omega) + \epsilon_2^2(\omega)} - \epsilon_1(\omega)} \tag{7}$$

The calculated absorption spectra, shown in Fig. 5(d), exhibit weak responses at low photon energies, followed by a rapid increase as optically allowed interband transitions become accessible. The strongest absorption occurs well above the fundamental electronic bandgaps. $Cs_2MoSeO_6$ reaches a maximum absorption coefficient of $2.037\times10^6$ $cm^{-1}$ at 16.55 eV, while $Rb_2MoSeO_6$ exhibits the highest overall value of $2.291\times10^6$ $cm^{-1}$ at 19.58 eV. $K_2MoSeO_6$ reaches a lower maximum of $1.367\times10^6$ $cm^{-1}$ at 24.29 eV.

The dominant absorption maximum therefore shifts toward lower photon energies from $K_2MoSeO_6$ to $Rb_2MoSeO_6$ and $Cs_2MoSeO_6$. This trend differs from the variation in the fundamental bandgaps, indicating that the strongest absorption features originate from transitions involving electronic states well beyond the immediate band edges. The substantial shift of the dominant high-energy peak, despite the smaller variation in the fundamental gap, illustrates the importance of transitions involving deeper valence and higher conduction states in controlling the extreme-ultraviolet response.

The optical penetration depth, defined as the distance over which intensity decreases to 1/e of its initial value, is given by

$$\delta(\omega) = \frac{1}{\alpha(\omega)} \tag{8}$$

At their respective absorption maxima, the penetration depths are approximately 7.3, 4.4, and 4.9 nm for $K_2MoSeO_6$, $Rb_2MoSeO_6$, and $Cs_2MoSeO_6$. These short penetration depths indicate strong attenuation of high-energy radiation within the idealized bulk materials.

The calculated absorption maxima lie predominantly in the extreme-ultraviolet region. Accordingly, the results are more directly relevant to high-energy optical attenuation and wavelength-selective optical response than to conventional visible-light photovoltaic absorption. The combination of large high-energy absorption coefficients and nanometer-scale penetration depths highlights the strong calculated attenuation response of these compositions in the extreme-ultraviolet range.

### 3.3.4. Reflectivity

The normal-incidence reflectivity at an ideal interface between air and a nonmagnetic material is obtained from the complex refractive index using the Fresnel relation [28], [29]:

$$R(\omega) = \left|\frac{N(\omega) - 1}{N(\omega) + 1}\right|^2 \tag{9}$$

Equivalently,

$$R(\omega) = \frac{[n(\omega) - 1]^2 + k^2(\omega)}{[n(\omega) + 1]^2 + k^2(\omega)} \tag{10}$$

The calculated low-energy reflectivities are 0.113, 0.133, and 0.129 for $K_2MoSeO_6$, $Rb_2MoSeO_6$, and $Cs_2MoSeO_6$, respectively. These values indicate modest reflection at an ideal air-material interface.

At higher energies, the reflectivity spectra in Fig. 5(e) become increasingly composition-dependent. $K_2MoSeO_6$ reaches a maximum reflectivity of 0.206 at 6.94 eV. In contrast, $Rb_2MoSeO_6$ and $Cs_2MoSeO_6$ exhibit stronger maxima of 0.336 at 20.27 eV and 0.339 at 17.18 eV, respectively. These enhanced high-energy reflectivity regions occur near the intervals where the real dielectric function becomes negative.

### 3.3.5. Electron Energy-Loss Function

The electron energy-loss function describes the energy dissipated by an energetic charged particle interacting with the electronic system of a material. It is obtained from the imaginary part of the negative inverse dielectric function [3], [12], [28]:

$$L(\omega) = Im\left[-\frac{1}{\epsilon(\omega)}\right] \quad (11)$$

which can be expressed as

$$L(\omega) = \frac{\epsilon_2(\omega)}{\epsilon_1^2(\omega) + \epsilon_2^2(\omega)} \quad (12)$$

The calculated energy-loss spectra in Fig. 5(f) exhibit relatively weak responses at low photon energies, followed by pronounced high-energy maxima. $K_2MoSeO_6$ shows a broad peak of 1.753 at 25.73 eV, while $Rb_2MoSeO_6$ reaches 3.602 at 24.84 eV. $Cs_2MoSeO_6$ exhibits the strongest and most sharply defined maximum, reaching 4.093 at 23.70 eV.

Photon energy corresponding to the peak of the loss function decreases systematically from $K_2MoSeO_6$ to $Cs_2MoSeO_6$, while the peak intensity increases. For $Rb_2MoSeO_6$ and $Cs_2MoSeO_6$, the real dielectric function approaches zero near the principal loss maxima, with relatively small imaginary components. The coincidence of a near-zero real dielectric response with comparatively weak damping is consistent with pronounced collective, plasmon-like electronic excitation features [3], [12], [28].

Notably, the energy-loss maxima do not coincide with the strongest optical absorption peaks. For example, $Cs_2MoSeO_6$ exhibits maximum absorption at 16.55 eV, whereas its principal energy-loss peak occurs at 23.70 eV. This distinction follows from the different mathematical dependence of the absorption coefficient and loss function on the dielectric response. The former describes electromagnetic attenuation, whereas the latter characterizes the response probed by charged-particle energy loss.

### 3.3.6. Optical Conductivity

The complex optical conductivity describes the electronic current induced by a time-varying electric field. Using the time dependence exp(−jωt), the optical conductivity is related to the dielectric function through [28], [29]

$$\sigma(\omega) = -j\omega\epsilon_0[\epsilon(\omega) - 1] \quad (13)$$

where ϵ0 is the vacuum permittivity. Separating the real and imaginary components gives

$$\sigma_1(\omega) = \omega\epsilon_0\epsilon_2(\omega) \quad (14)$$

and

$$\sigma_2(\omega) = \omega\epsilon_0[1 - \epsilon_1(\omega)] \quad (15)$$

The real optical conductivity represents the dissipative response associated with photon-induced electronic transitions, while the imaginary component describes the reactive polarization response. Equations (14) and (15) also explain the close correspondence between the optical conductivity and dielectric spectra.

As shown in Fig. 5(g), the real optical conductivity is negligible below the principal interband excitation region and increases as allowed electronic transitions become accessible. $K_2MoSeO_6$ reaches its maximum reported spectral value of 3.240 $fs^{-1}$ at 6.86 eV. $Rb_2MoSeO_6$ and $Cs_2MoSeO_6$ exhibit stronger high-energy maxima of 5.391 $fs^{-1}$ at 18.96 eV and 5.317 $fs^{-1}$ at 13.89 eV, respectively.

The different peak positions indicate that alkali-metal substitution redistributes the strengths of optically active electronic transitions. The correspondence between the real optical conductivity and imaginary dielectric function is expected from Eq. (12), because both originate from the same optically allowed interband excitations.

The imaginary optical conductivity, shown in Fig. 5(h), exhibits alternating positive and negative regions associated with the dispersive electronic response. Its largest positive spectral values occur at 25.13 eV for $K_2MoSeO_6$, 19.99 eV for $Rb_2MoSeO_6$, and 16.94 eV for $Cs_2MoSeO_6$, with reported magnitudes of 2.546, 4.213, and 3.748 $fs^{-1}$, respectively. The opposite signs of the reactive response over different spectral regions reflect changes in the phase relationship between the induced current and driving optical field, consistent with the dispersive dielectric spectra.

These spectral peaks quantify the frequency-dependent electronic excitation response and complement the dielectric and absorption results, rather than representing dc charge transport.

### 3.3.7. Comparative Optical Response

The calculated optical properties reveal distinct composition-dependent responses across the $X_2MoSeO_6$ (X = K, Rb, Cs) series. $Rb_2MoSeO_6$ exhibits the highest static dielectric constant, refractive index, and peak absorption coefficient, whereas $Cs_2MoSeO_6$ shows prominent absorption features at lower photon energies and $K_2MoSeO_6$ maintains a broader high-energy response. These variations, together with differences in the energy-loss spectra, reflect changes in interband transitions and electronic structure induced by alkali-metal substitution. The absence of a monotonic relationship between the bandgap and optical response highlights the influence of orbital contributions and transition strengths beyond the fundamental bandgap. Overall, these findings suggest opportunities for composition-tailored ultraviolet and extreme-ultraviolet optical applications, warranting further investigation of high-energy photon interactions.

## 3.4. Elastic Properties

The elastic properties of $X_2MoSeO_6$ (X = K, Rb, Cs) were examined to assess mechanical stability, resistance to deformation, and elastic anisotropy. The single-crystal elastic constants, $C_{ij}$, were calculated within the first-principles framework and used to determine the effective polycrystalline moduli through the Voigt-Reuss-Hill (VRH) approximation. The directional dependence of the elastic response was further investigated using ELATE [13], [14], [16].

### 3.4.1. Elastic Constants and Mechanical Stability

The second-order stiffness coefficients $C_{ij}$ (GPa) connect stress and elastic strain through generalized Hooke's law, $\sigma_i = \Sigma_j C_{ij}\varepsilon_j$, where $\sigma_i$ and $\varepsilon_j$ are the stress and strain components in Voigt notation, respectively. The diagonal longitudinal terms $C_{11}$, $C_{22}$, and $C_{33}$ describe resistance to normal strains along the crystallographic axes, while $C_{44}$, $C_{55}$, and $C_{66}$ characterize shear stiffness. Off-diagonal coefficients describe coupling between different deformation components.

The number of independent elastic constants depends on crystal symmetry. For monoclinic $K_2MoSeO_6$, the full stiffness tensor includes the symmetry-allowed coupling components $C_{15}$, $C_{25}$, $C_{35}$, and $C_{46}$. For orthorhombic $Rb_2MoSeO_6$ and $Cs_2MoSeO_6$, these components vanish by symmetry. Therefore, $K_2MoSeO_6$ has 13 independent elastic constants, whereas $Rb_2MoSeO_6$ and $Cs_2MoSeO_6$ have 9 independent constants (Table 3).

Table 3. Independent elastic stiffness constants $C_{ij}$ (GPa) for $X_2MoSeO_6$. Dashes denote components forbidden by orthorhombic symmetry.

| Compound | $C_{11}$ | $C_{22}$ | $C_{33}$ | $C_{44}$ | $C_{55}$ | $C_{66}$ | $C_{12}$ | $C_{13}$ | $C_{23}$ | $C_{15}$ | $C_{25}$ | $C_{35}$ | $C_{46}$ |
|---|---|---|---|---|---|---|---|---|---|---|---|---|---|
| $K_2MoSeO_6$ | 28.575 | 53.514 | 20.530 | 10.848 | 7.587 | 12.855 | 13.135 | 13.393 | 7.769 | 7.897 | −0.203 | 4.391 | −0.454 |
| $Rb_2MoSeO_6$ | 45.025 | 37.950 | 33.240 | 9.175 | 10.982 | 16.837 | 8.468 | 5.469 | 8.836 | — | — | — | — |
| $Cs_2MoSeO_6$ | 39.624 | 43.598 | 56.187 | 6.564 | 13.555 | 19.688 | 11.547 | 9.200 | 4.280 | — | — | — | — |

The calculated elastic constants in Table 3 reveal appreciable differences in directional stiffness. For $K_2MoSeO_6$, $C_{22}$ (53.514 GPa) is considerably larger than $C_{11}$ (28.575 GPa) and $C_{33}$ (20.530 GPa), indicating a pronounced directional contrast in the elastic response. Its nonzero coupling constants, particularly $C_{15} = 7.897$ GPa and $C_{35} = 4.391$ GPa, further reflect the lower symmetry of the monoclinic structure.

The longitudinal stiffness is greatest along the b axis for $K_2MoSeO_6$ ($C_{22} = 53.514$ GPa), whereas $Cs_2MoSeO_6$ exhibits its largest longitudinal coefficient along c ($C_{33} = 56.187$ GPa). These contrasts illustrate the different directional responses associated with their crystal structures.

For $Rb_2MoSeO_6$, the principal longitudinal stiffness constants follow $C_{11} > C_{22} > C_{33}$, with values of 45.025, 37.950, and 33.240 GPa, respectively. $Cs_2MoSeO_6$ exhibits the opposite ordering, $C_{33} > C_{22} > C_{11}$, with corresponding values of 56.187, 43.598, and 39.624 GPa. These differences indicate that alkali-metal substitution modifies the directional resistance to elastic deformation.

Mechanical stability under infinitesimal homogeneous deformation requires the elastic strain-energy density to be positive for every nonzero strain. This condition is satisfied when the symmetric elastic stiffness matrix is positive definite [13]. Equivalently, all eigenvalues of the stiffness matrix must be positive:

$$\lambda_i(C) > 0, i = 1, 2, \ldots, 6 \quad (16)$$

The six eigenvalues of each symmetric stiffness matrix are listed in Table 4. All are positive, confirming elastic stability against infinitesimal homogeneous strains for the investigated structures. This criterion does not, by itself, establish dynamical or finite-temperature stability.

Table 4. Eigenvalues (GPa) of the stiffness matrices

| Compound | $\lambda_1$ (GPa) | $\lambda_2$ (GPa) | $\lambda_3$ (GPa) | $\lambda_4$ (GPa) | $\lambda_5$ (GPa) | $\lambda_6$ (GPa) |
|---|---|---|---|---|---|---|
| $K_2MoSeO_6$ | 4.509 | 10.750 | 10.791 | 12.953 | 31.534 | 63.372 |
| $Rb_2MoSeO_6$ | 9.175 | 10.982 | 16.837 | 26.399 | 34.874 | 54.942 |
| $Cs_2MoSeO_6$ | 6.564 | 13.555 | 19.688 | 29.118 | 45.912 | 64.378 |

Table 5. Voigt, Reuss and Hill polycrystalline moduli, Poisson's ratio and Pugh ratio for $X_2MoSeO_6$.

| Compound | Scheme | B (GPa) | G (GPa) | E (GPa) | υ | B/G |
|---|---|---|---|---|---|---|
| $K_2MoSeO_6$ | Voigt | 19.024 | 10.813 | 27.272 | 0.261 | 1.761 |
| $K_2MoSeO_6$ | Reuss | 13.640 | 7.726 | 19.497 | 0.262 | 1.766 |
| $K_2MoSeO_6$ | Hill | 16.332 | 9.270 | 23.384 | 0.261 | 1.761 |
| $Rb_2MoSeO_6$ | Voigt | 17.973 | 13.628 | 32.636 | 0.197 | 1.319 |
| $Rb_2MoSeO_6$ | Reuss | 17.734 | 12.776 | 30.905 | 0.210 | 1.388 |
| $Rb_2MoSeO_6$ | Hill | 17.853 | 13.202 | 31.774 | 0.203 | 1.352 |
| $Cs_2MoSeO_6$ | Voigt | 21.051 | 15.587 | 37.504 | 0.203 | 1.350 |
| $Cs_2MoSeO_6$ | Reuss | 20.898 | 12.876 | 32.046 | 0.244 | 1.626 |
| $Cs_2MoSeO_6$ | Hill | 20.975 | 14.231 | 34.819 | 0.223 | 1.473 |

### 3.4.2. Polycrystalline Elastic Response

The bulk modulus B describes resistance to hydrostatic compression, while the shear modulus G characterizes resistance to shape change. Voigt and Reuss estimates provide upper and lower bounds for a randomly oriented polycrystalline aggregate; the Hill values are their arithmetic means. The elastic compliance matrix S is the inverse of the stiffness matrix C.

Young's modulus (E) and Poisson's ratio (ν) were obtained from the effective isotropic bulk and shear moduli. E quantifies uniaxial stiffness, whereas ν describes the associated transverse strain response.

$$B_V = \frac{C_{11} + C_{22} + C_{33} + 2(C_{12} + C_{13} + C_{23})}{9} \tag{17}$$

$$G_V = \frac{C_{11} + C_{22} + C_{33} - C_{12} - C_{13} - C_{23} + 3(C_{44} + C_{55} + C_{66})}{15} \tag{18}$$

$$B_R = \frac{1}{S_{11} + S_{22} + S_{33} + 2(S_{12} + S_{13} + S_{23})} \tag{19}$$

$$G_R = \frac{15}{4(S_{11} + S_{22} + S_{33}) - 4(S_{12} + S_{13} + S_{23}) + 3(S_{44} + S_{55} + S_{66})} \tag{20}$$

Here, $B_V$ and $B_R$ denote the Voigt and Reuss bulk moduli, respectively, while $G_V$ and $G_R$ are the corresponding shear moduli. The Hill averages and the corresponding isotropic Young's modulus and Poisson's ratio are obtained as follows:

$$B = \frac{B_V + B_R}{2}, \qquad G = \frac{G_V + G_R}{2} \tag{21}$$

$$E = \frac{9BG}{3B + G}, \qquad \nu = \frac{3B - 2G}{2(3B + G)} \tag{22}$$

The calculated Hill bulk moduli increase from 16.332 GPa for $K_2MoSeO_6$ to 17.853 GPa for $Rb_2MoSeO_6$ and 20.975 GPa for $Cs_2MoSeO_6$ (Table 5). The shear moduli follow the same trend, increasing from 9.270 to 13.202 and 14.231 GPa, respectively. Likewise, Young's modulus increases from 23.384 to 31.774 and 34.819 GPa. These results indicate a progressive increase in the average resistance to volumetric compression, shear deformation, and uniaxial elastic deformation across the K-Rb-Cs series. Although the three compounds exhibit relatively low absolute elastic moduli, their Hill-averaged resistance to volumetric, shear, and uniaxial deformation increases from $K_2MoSeO_6$ to $Cs_2MoSeO_6$.

The empirical Pugh ratio B/G [30] provides a qualitative indicator of ductile versus brittle tendencies. The Hill ratios of 1.761, 1.352, and 1.473 place $K_2MoSeO_6$ near the commonly used threshold of 1.75, while $Rb_2MoSeO_6$ and $Cs_2MoSeO_6$ fall on its brittle-associated side. Their corresponding isotropic Poisson's ratios (0.261, 0.203, and 0.223) support the same qualitative ordering; neither indicator directly determines fracture toughness.

### 3.4.3. Directional elastic anisotropy

Directional elastic anisotropy was evaluated using the Voigt–Reuss differences, the universal elastic anisotropy index ($A^U$), and the direction-dependent elastic properties obtained using

ELATE [15], [16] [15], [16]. The index $A^U$ provides a dimensionless measure of the departure from elastic isotropy and is defined as follows,

$$A^U = 5\frac{G_V}{G_R} + \frac{B_V}{B_R} - 6 \tag{23}$$

For an isotropic material, $A^U = 0$; larger positive values indicate stronger elastic anisotropy. The Voigt–Reuss bulk-modulus differences are 5.384, 0.239, and 0.153 GPa for $K_2MoSeO_6$, $Rb_2MoSeO_6$, and $Cs_2MoSeO_6$, respectively; the corresponding shear-modulus differences are 3.087, 0.852, and 2.711 GPa (Table 5). Using Eq. (23), the calculated $A^U$ values are approximately 2.39, 0.35, and 1.06, respectively.

The ELATE analysis resolves Young's modulus, shear modulus, linear compressibility, and Poisson's ratio as functions of direction (Figs. 6-8). Linear compressibility is the fractional change in length along a specified direction under hydrostatic pressure; its extrema are reported alongside the other directional elastic properties in Tables 6-8.

Table 6. Directional elastic extrema and anisotropy ratios from ELATE for $K_2MoSeO_6$.

| Parameters | Young's Modulus (GPa) | Young's Modulus (GPa) | Lin. Compressibility (TPa−1) | Lin. Compressibility (TPa−1) | Shear Modulus (GPa) | Shear Modulus (GPa) | Poisson's Ratio | Poisson's Ratio |
|---|---|---|---|---|---|---|---|---|
| | $E_{min}$ | $E_{max}$ | $\beta_{min}$ | $\beta_{max}$ | $G_{min}$ | $G_{max}$ | $\upsilon_{min}$ | $\upsilon_{max}$ |
| Values | 9.084 | 44.484 | 5.854 | 60.276 | 4.101 | 17.362 | −0.107 | 0.793 |
| Anisotropy | 4.897 | 4.897 | 10.296 | 10.296 | 4.233 | 4.233 | ∞ | ∞ |

Table 7. Directional elastic extrema and anisotropy ratios from ELATE for $Rb_2MoSeO_6$.

| Parameters | Young's Modulus (GPa) | Young's Modulus (GPa) | Lin. Compressibility (TPa−1) | Lin. Compressibility (TPa−1) | Shear Modulus (GPa) | Shear Modulus (GPa) | Poisson's Ratio | Poisson's Ratio |
|---|---|---|---|---|---|---|---|---|
| | $E_{min}$ | $E_{max}$ | $\beta_{min}$ | $\beta_{max}$ | $G_{min}$ | $G_{max}$ | $\upsilon_{min}$ | $\upsilon_{max}$ |
| Values | 25.511 | 42.743 | 16.162 | 22.790 | 9.175 | 16.836 | 0.081 | 0.394 |
| Anisotropy | 1.675 | 1.675 | 1.410 | 1.410 | 1.835 | 1.835 | 4.858 | 4.858 |

Table 8. Directional elastic extrema and anisotropy ratios from ELATE for $Cs_2MoSeO_6$.

| Parameters | Young's Modulus (GPa) | Young's Modulus (GPa) | Lin. Compressibility (TPa−1) | Lin. Compressibility (TPa−1) | Shear Modulus (GPa) | Shear Modulus (GPa) | Poisson's Ratio | Poisson's Ratio |
|---|---|---|---|---|---|---|---|---|
| | $E_{min}$ | $E_{max}$ | $\beta_{min}$ | $\beta_{max}$ | $G_{min}$ | $G_{max}$ | $\upsilon_{min}$ | $\upsilon_{max}$ |
| Values | 20.550 | 53.987 | 13.701 | 17.083 | 6.564 | 22.598 | −0.038 | 0.586 |
| Anisotropy | 2.627 | 2.627 | 1.247 | 1.247 | 3.443 | 3.443 | ∞ | ∞ |

For $K_2MoSeO_6$, the directional Young's modulus ranges from 9.084 to 44.484 GPa ($E_{max}/E_{min}$ = 4.897). Linear compressibility spans 5.854–60.276 $TPa^{-1}$, while the shear modulus varies from 4.101 to 17.362 GPa. These broad ranges indicate strong orientation dependence.

$Rb_2MoSeO_6$ exhibits a narrower directional Young's modulus range of 25.511–42.743 GPa ($E_{max}/E_{min}$ = 1.675). Its linear compressibility spans 16.162–22.790 $TPa^{-1}$, and its shear modulus varies from 9.175 to 16.836 GPa. These comparatively small directional variations are consistent with its low universal anisotropy index.

$Cs_2MoSeO_6$ exhibits intermediate anisotropy in Young's modulus (20.550–53.987 GPa; $E_{max}/E_{min}$ = 2.627). Its linear compressibility varies from 13.701 to 17.083 $TPa^{-1}$, whereas its shear modulus spans 6.564–22.598 GPa, revealing pronounced shear anisotropy despite a relatively narrow compressibility range.

The directional Poisson's ratios reveal negative minima for $K_2MoSeO_6$ (−0.107) and $Cs_2MoSeO_6$ (−0.038), whereas $Rb_2MoSeO_6$ remains positive over the evaluated directions (0.081–0.394). The negative values indicate an auxetic response—lateral expansion under uniaxial tension—for particular loading and transverse directions.

As evident from the anisotropy indices and the directional ELATE results, $K_2MoSeO_6$ exhibits the strongest overall elastic anisotropy and the lowest minimum directional Young's modulus (9.084 GPa). $Rb_2MoSeO_6$ is the most nearly isotropic, whereas $Cs_2MoSeO_6$ combines the highest Hill-averaged moduli with appreciable shear anisotropy.

In summary, all three investigated structures satisfy the elastic stability criterion, and their Hill-averaged moduli increase across the K-Rb-Cs series. The marked differences in anisotropy and directional Poisson's ratio show that alkali-metal substitution changes not only average stiffness but also the orientation dependence of elastic deformation.

### 3.5. Thermoelectric Properties

The thermoelectric response of $K_2MoSeO_6$, $Rb_2MoSeO_6$, and $Cs_2MoSeO_6$ was investigated from 100 to 800 K using semiclassical Boltzmann transport theory within the constant relaxation-time approximation (CRTA) [20], [31] (Fig. 9). Electronic eigenvalues obtained from density functional theory calculations using Quantum ESPRESSO [22] were processed with BoltzTraP2 [20] to determine the Seebeck coefficient (S), electrical conductivity normalized by relaxation time (σ/τ), electronic thermal conductivity normalized by relaxation time ($\kappa_e/\tau$), power factor normalized by relaxation time (PF/τ), and electronic-only figure of merit ($ZT_e$). This framework enables a systematic comparison of carrier-polarity-dependent transport across the K-Rb-Cs series.

For each compound, the chemical-potential dependence of S was examined at 300 K. The chemical potentials corresponding to the largest positive and most negative Seebeck coefficients on the sampled grid were selected to represent hole-dominated and electron-dominated transport, respectively [32]–[34]. The resulting pairs were 3.8030/6.5991 eV for $K_2MoSeO_6$, 2.9994/4.8004 eV for $Rb_2MoSeO_6$, and 5.2635/7.3315 eV for $Cs_2MoSeO_6$ (positive-/negative-Seebeck conditions). Holding these chemical potentials fixed over 100-800 K isolates how the transport response evolves with temperature under each selected occupation condition. The selection is designed to compare thermopower extrema, rather than to optimize PF or $ZT_e$ directly.

### 3.5.1. Transport Formalism

Within the semiclassical Boltzmann framework, thermoelectric transport coefficients are determined by the distribution of electronic states, their group velocities, and their occupation probabilities. The Fermi-Dirac distribution is expressed as

$$f(E, \mu, T) = \frac{1}{1 + \exp\left(\frac{E - \mu}{k_B T}\right)} \quad (24)$$

where E denotes electron energy, μ is the chemical potential, $k_B$ is Boltzmann's constant, and T is the absolute temperature.

The transport moments are defined as

$$\mathcal{L}_n(\mu, T) = \int_{-\infty}^{\infty} \Xi(E)\ (E - \mu)^n \left(-\frac{\partial f}{\partial E}\right)\ dE \quad (25)$$

where Ξ(E) represents the energy-dependent transport distribution function excluding the constant relaxation time. This function incorporates the contributions of electronic band velocities and the available electronic states. The derivative of the Fermi-Dirac distribution acts as a weighting function that selects states within a thermal energy window around the chemical potential.

The Seebeck coefficient is obtained from the first and zeroth transport moments as

$$S(\mu, T) = -\frac{1}{eT}\frac{\mathcal{L}_1(\mu, T)}{\mathcal{L}_0(\mu, T)} \quad (26)$$

where e is the positive elementary charge. Equation (26) establishes the direct relationship between chemical potential and thermopower. Electronic states above and below μ contribute with opposite signs to the first transport moment. Consequently, electron-dominated transport produces a negative Seebeck coefficient, whereas hole-dominated transport produces a positive value.

For a sufficiently smooth transport distribution in the low-temperature, degenerate limit, the Mott relation provides a useful approximation [35], [36]:

$$S = -\frac{\pi^2\, k_B^2\, T}{3e} \left.\frac{d\ln\Xi(E)}{dE}\right|_{E=\mu} \tag{27}$$

This expression shows that thermopower is governed by the energy dependence of the transport distribution near the chemical potential rather than simply by the density of electronic states. A steep positive slope of Ξ(E) near μ favors negative thermopower, while a negative slope favors positive thermopower. The Mott relation is used here to explain the physical trends; the calculated transport coefficients are based on the full Boltzmann transport expressions.

### 3.5.2. Seebeck Coefficient

The calculated Seebeck coefficients exhibit a pronounced dependence on chemical potential, with a reversal in sign between the two selected conditions for all three compounds (Fig. 9a). Positive values are obtained at μ = 3.8030, 2.9994, and 5.2635 eV for $K_2MoSeO_6$, $Rb_2MoSeO_6$, and $Cs_2MoSeO_6$, respectively, indicating predominantly hole-like transport. At the second chemical potentials of 6.5991, 4.8004, and 7.3315 eV, the coefficients become negative, consistent with electron-like transport.

Among the positive-Seebeck cases, $Rb_2MoSeO_6$ exhibits the largest thermopower, reaching 390.55 μV K$^{-1}$ at 100 K for μ = 2.9994 eV, before decreasing to 88.86 μV K$^{-1}$ at 800 K. $K_2MoSeO_6$ displays a more moderate response, reaching 161.01 μV K$^{-1}$ at 200 K and subsequently decreasing to 118.65 μV K$^{-1}$ at 800 K. $Cs_2MoSeO_6$ exhibits considerably smaller positive thermopower, with a maximum of approximately 47.00 μV K$^{-1}$ near 300 K.

In the negative-Seebeck regime, $K_2MoSeO_6$ exhibits the largest magnitude at low temperature, reaching −299.77 μV K$^{-1}$ at 100 K. $Cs_2MoSeO_6$ attains approximately −99.65 μV K$^{-1}$ near 250 K, whereas the negative thermopower of $Rb_2MoSeO_6$ remains comparatively small throughout the investigated temperature range.

The pronounced positive thermopower of $Rb_2MoSeO_6$ and negative thermopower of $K_2MoSeO_6$ point to substantial energy asymmetry in the transport distribution near their selected chemical potentials. Besides, the different temperature profiles reflect composition-dependent electronic dispersion and the energy selectivity of charge transport. As temperature rises, broadening of the Fermi-Dirac weighting window redistributes the relative contributions of states above and below μ, explaining why the magnitude of S decreases in several cases [19], [20].

### 3.5.3. Electrical Conductivity

The electrical conductivity per relaxation time is expressed in terms of the zeroth transport moment as

$$\frac{\sigma}{\tau} = e^2 \mathcal{L}_0(\mu, T) \tag{28}$$

Here, τ denotes the carrier relaxation time (in seconds), while σ/τ represents the electrical conductivity normalized by τ, as calculated from the electronic band structure within the constant relaxation-time approximation.

For all three compounds, σ/τ increases with temperature under the chemical potentials corresponding to positive Seebeck coefficients (Fig. 9b). In particular, $K_2MoSeO_6$ exhibits increasing conductivity under both selected chemical potentials, reaching $2.56 \times 10^{18}$ and $2.31 \times 10^{18}$ $(\Omega\ \mathrm{m\ s})^{-1}$ at 800 K. The common increasing trend in the positive-Seebeck regimes suggests that thermal broadening of the Fermi-Dirac distribution progressively enhances the contribution of electronic states with relatively large transport weights. Physically, these states may include valence-band states with higher group velocities or additional thermally accessible bands that contribute to the conductivity integral. Because the selected chemical potentials remain fixed, increasing temperature changes the range and relative weighting of electronic states participating in transport, producing a net increase in σ/τ [20], [31].

In contrast, the negative-Seebeck regimes reveal material-dependent conductivity responses. $Rb_2MoSeO_6$ exhibits a decrease in σ/τ from $1.62 \times 10^{18}$ to $0.95 \times 10^{18}$ $(\Omega\ \mathrm{m\ s})^{-1}$ between 100 and 800 K at μ = 4.8004 eV. This behavior is consistent with a transport distribution in which the states initially sampled near the selected chemical potential contribute more strongly to conductivity than the additional states incorporated as the thermal occupation window broadens. Consequently, the weighted average transport contribution decreases with increasing temperature. A similar decreasing trend is observed for $Cs_2MoSeO_6$ under its negative-Seebeck condition (μ = 7.3315 eV), suggesting that its electron-like transport distribution exhibits a comparable local energy dependence near the selected chemical potential.

Interestingly, $K_2MoSeO_6$ maintains an increasing σ/τ even under its negative-Seebeck condition (μ = 6.5991 eV), distinguishing it from $Rb_2MoSeO_6$ and $Cs_2MoSeO_6$. This contrasting response indicates that the transport distribution of $K_2MoSeO_6$ has a different energy dependence around the selected electron-like chemical potential. As temperature increases, the additional thermally sampled electronic states appear to provide a net enhancement of the conductivity integral rather than reducing its weighted value. Such differences may arise from variations in band dispersion, electronic group velocities, and the availability of transport-active states near the relevant chemical potentials [20], [31].

The contrasting σ/τ trends arise from the interplay between the energy-resolved transport distribution and thermal occupation at fixed μ. In particular, the declining electron-like σ/τ of $Rb_2MoSeO_6$ and $Cs_2MoSeO_6$ reflects reduced weighting of the most conductive states as the Fermi window broadens, whereas $K_2MoSeO_6$ retains a net increase from newly sampled transport-active states. These are band-occupation effects within CRTA, rather than changes in phonon-limited scattering time [20], [31].

### 3.5.4. Electronic Thermal Conductivity

The electronic thermal conductivity per relaxation time is obtained from the transport moments as

$$\frac{\kappa_e}{\tau} = \frac{1}{T}\left[\mathcal{L}_2(\mu, T) - \frac{\mathcal{L}_1^2(\mu, T)}{\mathcal{L}_0(\mu, T)}\right] \tag{29}$$

The first term describes the energy carried by electronic excitations, while the second accounts for thermoelectric coupling under zero electrical current. Here $\kappa_e$ denotes the electronic contribution to thermal conductivity; the phonon-mediated lattice contribution κl is not included.

The calculated electronic thermal conductivity generally increases with temperature (Fig. 9c), indicating enhanced electronic heat transport as a wider range of electronic states becomes thermally accessible. This increase is particularly pronounced for $Cs_2MoSeO_6$ at $\mu = 5.2635$ eV, where $\kappa_e/\tau$ reaches $7.74 \times 10^{13}$ W $m^{-1}$ $K^{-1}$ $s^{-1}$ at 800 K. $K_2MoSeO_6$ also exhibits increasing electronic thermal conductivity under both chemical-potential conditions.

Electrical and electronic thermal transport are linked through the same electronic states but involve different energy weightings. Accordingly, the particularly strong rise in $\kappa_e/\tau$ for $Cs_2MoSeO_6$ demonstrates that enhanced electronic conduction can be accompanied by substantial electronic heat flow. This interplay provides a physical basis for comparing PF/τ and $ZT_e$ rather than evaluating σ/τ alone.

### 3.5.5. Power Factor and Thermoelectric Figure of Merit

The thermoelectric power factor is defined as

$$PF = S^2\,\sigma \tag{30}$$

and, within the constant relaxation-time approximation, can be written as

$$\frac{PF}{\tau} = S^2\frac{\sigma}{\tau} \tag{31}$$

The power factor represents the competition between thermopower and electrical conduction. A large Seebeck coefficient is desirable, but its benefit may be offset by low electrical conductivity. This trade-off is evident in $Rb_2MoSeO_6$ at $\mu = 2.9994$ eV (Fig. 9d), where the exceptionally large low-temperature thermopower is accompanied by comparatively weak conductivity. Conversely, $Cs_2MoSeO_6$ exhibits relatively high conductivity at $\mu = 5.2635$ eV but modest thermopower. $K_2MoSeO_6$ provides a more balanced combination of the two quantities under the selected chemical potentials.

The complete dimensionless thermoelectric figure of merit is defined as

$$ZT = \frac{S^2 \sigma T}{\kappa_e + \kappa_l} \tag{32}$$

where $\kappa_l$ represents lattice thermal conductivity. Since lattice thermal conductivity was not included in the present calculations, the reported values correspond to the electronic-only figure of merit, ZTe:

$$ZT_e = \frac{S^2 \sigma T}{\kappa_e} = \frac{S^2 (\sigma/\tau) T}{(\kappa_e/\tau)} \tag{33}$$

Substituting Eqs. (26), (28), and (29) into Eq. (33) gives

$$ZT_e = \frac{\mathcal{L}_1^2}{\mathcal{L}_0 \mathcal{L}_2 - \mathcal{L}_1^2} \tag{34}$$

At 300 K, $K_2MoSeO_6$ exhibits $ZT_e$ values of 2.410 and 3.662 at $\mu$ = 3.8030 and 6.5991 eV, respectively (Fig. 9e). The corresponding values for $Rb_2MoSeO_6$ are 7.286 and 0.083, while $Cs_2MoSeO_6$ yields 0.092 and 1.196. The substantial differences between the two chemical potentials demonstrate the sensitivity of electronic thermoelectric performance to carrier occupation.

$Rb_2MoSeO_6$ achieves the largest electronic-only figure of merit, $ZT_e$ = 28.83 at 100 K under its positive-Seebeck condition, reflecting a particularly favorable ratio of thermopower-generated electrical response to electronic heat conduction. This result identifies a pronounced electronic transport advantage at the selected chemical potential. Because $ZT_e$ is defined using $\kappa_e$ alone, it is distinct from the full device-relevant ZT, which additionally includes lattice thermal conductivity.

$K_2MoSeO_6$ exhibits comparatively favorable electronic-only performance across both chemical potentials. Under the negative-Seebeck condition, $ZT_e$ decreases from 7.332 at 100 K to 2.657 at 800 K. $Cs_2MoSeO_6$ reaches its largest $ZT_e$ of approximately 1.746 near 550 K at $\mu$ = 7.3315 eV, whereas its positive-Seebeck condition produces considerably smaller values.

### 3.5.6. Overall Thermoelectric Response and Application Prospects

Taken together, the calculations reveal distinct composition- and chemical-potential-dependent thermoelectric responses among the investigated compounds. $Rb_2MoSeO_6$ exhibits strong positive thermopower and a high electronic-only figure of merit ($ZT_e$) under the selected hole-like transport condition, whereas $K_2MoSeO_6$ demonstrates substantial negative thermopower under the selected electron-like condition. $Cs_2MoSeO_6$ exhibits comparatively strong electrical and electronic thermal transport, particularly in the positive-Seebeck regime. These contrasting characteristics motivate further investigation of carrier-concentration optimization for thermoelectric sensing and low-grade or moderate-temperature heat harvesting [37]–[39]. However, assessing their practical energy-conversion potential requires consideration of the lattice thermal conductivity and the resulting total thermoelectric figure of merit (ZT).

## 4. Conclusions

A comparative first-principles investigation of $X_2MoSeO_6$ (X = K, Rb, and Cs) has revealed the pronounced influence of alkali-metal substitution on the structural, electronic, elastic, optical, and thermoelectric properties of these compounds. The optimized lattice parameters of $K_2MoSeO_6$ and $Rb_2MoSeO_6$ agree well with available experimental crystallographic data, while the investigated $Cs_2MoSeO_6$ structure adopts an orthorhombic *Pnma* configuration. The unit-cell volume increases systematically from 663.162 to 791.034 Å$^3$ across the K-Rb-Cs series, whereas the electronic bandgap exhibits a distinctly non-monotonic variation. All three compounds display indirect semiconducting gaps of 2.646, 2.216, and 2.427 eV, respectively, with the valence-band edges dominated by O-p states and the lower conduction bands primarily associated with Mo-d orbitals. This contrast between progressive lattice expansion and non-monotonic bandgap evolution demonstrates that electronic behavior cannot be inferred from changes in unit-cell dimensions alone.

The calculated elastic stiffness tensors satisfy the mechanical stability criteria for all three investigated structures. The Hill-averaged bulk and shear moduli increase from 16.332 and 9.270 GPa for $K_2MoSeO_6$ to 20.975 and 14.231 GPa for $Cs_2MoSeO_6$, respectively, indicating greater resistance to volumetric and shear deformation in the Cs-containing compound. Nevertheless, elastic anisotropy follows a different trend, with $K_2MoSeO_6$ exhibiting the largest universal anisotropy index (2.39), compared with 0.35 and 1.06 for $Rb_2MoSeO_6$ and $Cs_2MoSeO_6$. The optical calculations reveal further composition-dependent differences: $Rb_2MoSeO_6$ possesses the highest static electronic dielectric constant (4.614) and maximum absorption coefficient ($2.291 \times 10^6$ cm$^{-1}$), whereas $Cs_2MoSeO_6$ exhibits the strongest principal electron energy-loss peak. These results establish that mechanical stiffness, elastic anisotropy, and high-energy optical response evolve differently with alkali-metal substitution, reflecting the distinct sensitivities of these properties to crystal structure and electronic excitations.

The thermoelectric analysis further demonstrates the importance of carrier occupation in determining transport behavior. $Rb_2MoSeO_6$ exhibits the largest positive Seebeck coefficient of 390.55 μV K$^{-1}$ at 100 K under the selected hole-like condition, while $K_2MoSeO_6$ reaches a negative Seebeck coefficient of −299.77 μV K$^{-1}$ at the same temperature under its selected electron-like condition. At 300 K, the highest electronic-only thermoelectric figure of merit among the investigated conditions is obtained for $Rb_2MoSeO_6$ ($ZT_e = 7.286$) in the positive-Seebeck regime. These values characterize electronic transport within the constant relaxation-time approximation; assessment of the complete thermoelectric performance additionally requires lattice thermal conductivity. Overall, this study extends the understanding of the heavier alkali-metal members of the $X_2MoSeO_6$ family by establishing composition-dependent relationships among crystal symmetry, electronic structure, mechanical response, optical excitations, and carrier-dependent transport. The findings identify distinct properties of the K-, Rb-, and Cs-containing compounds and provide a basis for targeted investigations of their optical and thermoelectric functionalities.

## References

[1] R. D. Shannon, “Revised effective ionic radii and systematic studies of interatomic

distances in halides and chalcogenides," *Acta Crystallogr. Sect. A*, vol. 32, no. 5, pp. 751–767, Sep. 1976, doi: 10.1107/S0567739476001551.

[2] J. P. Perdew *et al.*, "Restoring the Density-Gradient Expansion for Exchange in Solids and Surfaces," *Phys. Rev. Lett.*, vol. 100, no. 13, p. 136406, Apr. 2008, doi: 10.1103/PhysRevLett.100.136406.

[3] G. Onida, L. Reining, and A. Rubio, "Electronic excitations: density-functional versus many-body Green's-function approaches," *Rev. Mod. Phys.*, vol. 74, no. 2, pp. 601–659, Jun. 2002, doi: 10.1103/RevModPhys.74.601.

[4] Y. Porter and P. S. Halasyamani, "New alkali–metal–molybdenum(VI)–selenium(IV) oxides: syntheses, structures, and characterization of A2SeMoO6 (A=Na+, K+, or Rb+)," *J. Solid State Chem.*, vol. 174, no. 2, pp. 441–449, Sep. 2003, doi: 10.1016/S0022-4596(03)00296-2.

[5] A. Khan, N. Ullah Khan, A. Nawaz, K. Ullah, and A. Manan, "A DFT study to explore structural, electronic, optical and mechanical properties of lead-free Na2MoXO6 (X = Si, Ge, Sn) double perovskites for photovoltaic and optoelectronic applications," *Comput. Theor. Chem.*, vol. 1240, p. 114834, Oct. 2024, doi: 10.1016/j.comptc.2024.114834.

[6] J. Li, L. Sun, X. Cao, and J. Chang, "First-Principles Predictions of Structural, Mechanical, and Optoelectronic Properties of Se-Based Double Perovskites A 2 SeX 6 (A = Rb, K; X = Cl, Br, I)," *J. Phys. Chem. C*, vol. 127, no. 21, pp. 10332–10340, Jun. 2023, doi: 10.1021/acs.jpcc.3c01097.

[7] M. Manzoor *et al.*, "A DFT study: Tailoring opto-electronic and thermoelectric performance of K2SeX6 (X = Cl, Br) double perovskites for solar cell advancements," *Chinese J. Phys.*, vol. 89, pp. 278–289, Jun. 2024, doi: 10.1016/j.cjph.2023.09.016.

[8] H. Zhang *et al.*, "The impact of halogens on the structural, electronic, and optical properties of vacancy-ordered double perovskites Rb2SeX6 (X=I, Br, Cl)," *J. Solid State Chem.*, vol. 340, p. 125002, Dec. 2024, doi: 10.1016/j.jssc.2024.125002.

[9] S. Zahan, D. Afroj, and M. Abdur Rashid, "Impact of pressure on structural, mechanical, optoelectronic and thermoelectric properties of vacancy-ordered double perovskite K <math altimg="si97.svg" display="inline" id="d1e1500"> <msub> <mrow/> <mrow> <mn>2</mn> </mrow> </msub> </math> SeCl <math al," *Next Mater.*, vol. 8, p. 100512, Jul. 2025, doi: 10.1016/j.nxmate.2025.100512.

[10] S. Abdalla *et al.*, "Probing the Physical Attributes of Double Perovskites X2MoSeO6 (X = Na, Li) Through Advanced Computational Methods," *J. Inorg. Organomet. Polym. Mater.*, vol. 35, no. 10, pp. 8310–8323, Oct. 2025, doi: 10.1007/s10904-025-03818-0.

[11] D. R. Penn, "Wave-Number-Dependent Dielectric Function of Semiconductors," *Phys. Rev.*, vol. 128, no. 5, pp. 2093–2097, Dec. 1962, doi: 10.1103/PhysRev.128.2093.

[12] S. L. Adler, "Quantum Theory of the Dielectric Constant in Real Solids," *Phys. Rev.*, vol. 126, no. 2, pp. 413–420, Apr. 1962, doi: 10.1103/PhysRev.126.413.

[13] F. Mouhat and F.-X. Coudert, "Necessary and sufficient elastic stability conditions in various crystal systems," *Phys. Rev. B*, vol. 90, no. 22, p. 224104, Dec. 2014, doi:

10.1103/PhysRevB.90.224104.

[14] R. Hill, “The Elastic Behaviour of a Crystalline Aggregate,” *Proc. Phys. Soc. Sect. A*, vol. 65, no. 5, pp. 349–354, May 1952, doi: 10.1088/0370-1298/65/5/307.

[15] S. I. Ranganathan and M. Ostoja-Starzewski, “Universal Elastic Anisotropy Index,” *Phys. Rev. Lett.*, vol. 101, no. 5, p. 055504, Aug. 2008, doi: 10.1103/PhysRevLett.101.055504.

[16] R. Gaillac, P. Pullumbi, and F.-X. Coudert, “ELATE: an open-source online application for analysis and visualization of elastic tensors,” *J. Phys. Condens. Matter*, vol. 28, no. 27, p. 275201, Jul. 2016, doi: 10.1088/0953-8984/28/27/275201.

[17] G. J. Snyder and E. S. Toberer, “Complex thermoelectric materials,” *Nat. Mater.*, vol. 7, no. 2, pp. 105–114, Feb. 2008, doi: 10.1038/nmat2090.

[18] J. He and T. M. Tritt, “Advances in thermoelectric materials research: Looking back and moving forward,” *Science (80-. ).*, vol. 357, no. 6358, Sep. 2017, doi: 10.1126/science.aak9997.

[19] G. D. Mahan and J. O. Sofo, “The best thermoelectric.,” *Proc. Natl. Acad. Sci.*, vol. 93, no. 15, pp. 7436–7439, Jul. 1996, doi: 10.1073/pnas.93.15.7436.

[20] G. K. H. Madsen, J. Carrete, and M. J. Verstraete, “BoltzTraP2, a program for interpolating band structures and calculating semi-classical transport coefficients,” *Comput. Phys. Commun.*, vol. 231, pp. 140–145, Oct. 2018, doi: 10.1016/j.cpc.2018.05.010.

[21] S. J. Clark *et al.*, “First principles methods using CASTEP,” *Zeitschrift für Krist. - Cryst. Mater.*, vol. 220, no. 5–6, pp. 567–570, May 2005, doi: 10.1524/zkri.220.5.567.65075.

[22] P. Giannozzi *et al.*, “QUANTUM ESPRESSO: a modular and open-source software project for quantum simulations of materials,” *J. Phys. Condens. Matter*, vol. 21, no. 39, p. 395502, Sep. 2009, doi: 10.1088/0953-8984/21/39/395502.

[23] P. Giannozzi *et al.*, “Advanced capabilities for materials modelling with Q <scp>uantum</scp> ESPRESSO,” *J. Phys. Condens. Matter*, vol. 29, no. 46, p. 465901, Nov. 2017, doi: 10.1088/1361-648X/aa8f79.

[24] H. J. Monkhorst and J. D. Pack, “Special points for Brillouin-zone integrations,” *Phys. Rev. B*, vol. 13, no. 12, pp. 5188–5192, Jun. 1976, doi: 10.1103/PhysRevB.13.5188.

[25] D. D. Koelling and B. N. Harmon, “A technique for relativistic spin-polarised calculations,” *J. Phys. C Solid State Phys.*, vol. 10, no. 16, pp. 3107–3114, Aug. 1977, doi: 10.1088/0022-3719/10/16/019.

[26] C. M. YU P, *Fundamentals of semiconductors: physics and materials properties*. Berlin Heidelberg: Springer-Verlag, 2010.

[27] M. Zacharias, C. E. Patrick, and F. Giustino, “Stochastic Approach to Phonon-Assisted Optical Absorption,” *Phys. Rev. Lett.*, vol. 115, no. 17, p. 177401, Oct. 2015, doi: 10.1103/PhysRevLett.115.177401.

[28] F. Wooten, Frederick and Stern, *Optical properties of solids*. American Institute of

Physics, 1973.

[29] M. Fox, *Optical properties of solids*. Oxford university press., 2010.

[30] S. F. Pugh, "XCII. Relations between the elastic moduli and the plastic properties of polycrystalline pure metals," *London, Edinburgh, Dublin Philos. Mag. J. Sci.*, vol. 45, no. 367, pp. 823–843, Aug. 1954, doi: 10.1080/14786440808520496.

[31] G. K. H. Madsen and D. J. Singh, "BoltzTraP. A code for calculating band-structure dependent quantities," *Comput. Phys. Commun.*, vol. 175, no. 1, pp. 67–71, Jul. 2006, doi: 10.1016/j.cpc.2006.03.007.

[32] M. Yazdani-Kachoei and S. Jalali-Asadabadi, "Thermoelectric properties of heavy fermion CeRhIn 5 using density functional theory combined with semiclassical Boltzmann theory," *RSC Adv.*, vol. 9, no. 62, pp. 36182–36197, 2019, doi: 10.1039/C9RA07859B.

[33] E. Pakizeh, J. Jalilian, and M. Mohammadi, "Electronic, optical and thermoelectric properties of Fe 2 ZrP compound determined via first-principles calculations," *RSC Adv.*, vol. 9, no. 44, pp. 25900–25911, 2019, doi: 10.1039/C9RA04736K.

[34] M. Yazdani-Kachoei, S. Rahimi, R. Ebrahimi-Jaberi, J. Nematollahi, and S. Jalali-Asadabadi, "Thermoelectric properties plus phonon and de Haas–van Alphen frequencies of hole/electron-doped $$\hbox {CeIn}_3$$," *Sci. Rep.*, vol. 12, no. 1, p. 663, Jan. 2022, doi: 10.1038/s41598-021-04058-1.

[35] H. NF Mott, Sir and Jones, *The theory of the properties of metals and alloys*. Oxford, Clarendon Press, 1936.

[36] G. B. Ziman, John M and Ziman, John Michael and Ziman, John Michael and Physicien, Grande-Bretagne and Ziman, John Michael and Physicist, *Electrons and phonons: the theory of transport phenomena in solids*. Clarendon press, Oxford, 1960.

[37] J. Liu, Q. Liu, S. Lin, M. Y. Leung, Y. Ma, and X. Tao, "Wearable Thermoelectric Generators: Materials, Structures, Fabrications, and Applications," *Phys. status solidi – Rapid Res. Lett.*, vol. 17, no. 7, Jul. 2023, doi: 10.1002/pssr.202200502.

[38] Z. Tabaie and A. Omidvar, "Human body heat-driven thermoelectric generators as a sustainable power supply for wearable electronic devices: Recent advances, challenges, and future perspectives," *Heliyon*, vol. 9, no. 4, p. e14707, Apr. 2023, doi: 10.1016/j.heliyon.2023.e14707.

[39] K.-T. Lee *et al.*, "An overview of commercialization and marketization of thermoelectric generators for low-temperature waste heat recovery," *iScience*, vol. 26, no. 10, p. 107874, Oct. 2023, doi: 10.1016/j.isci.2023.107874.

## Figures

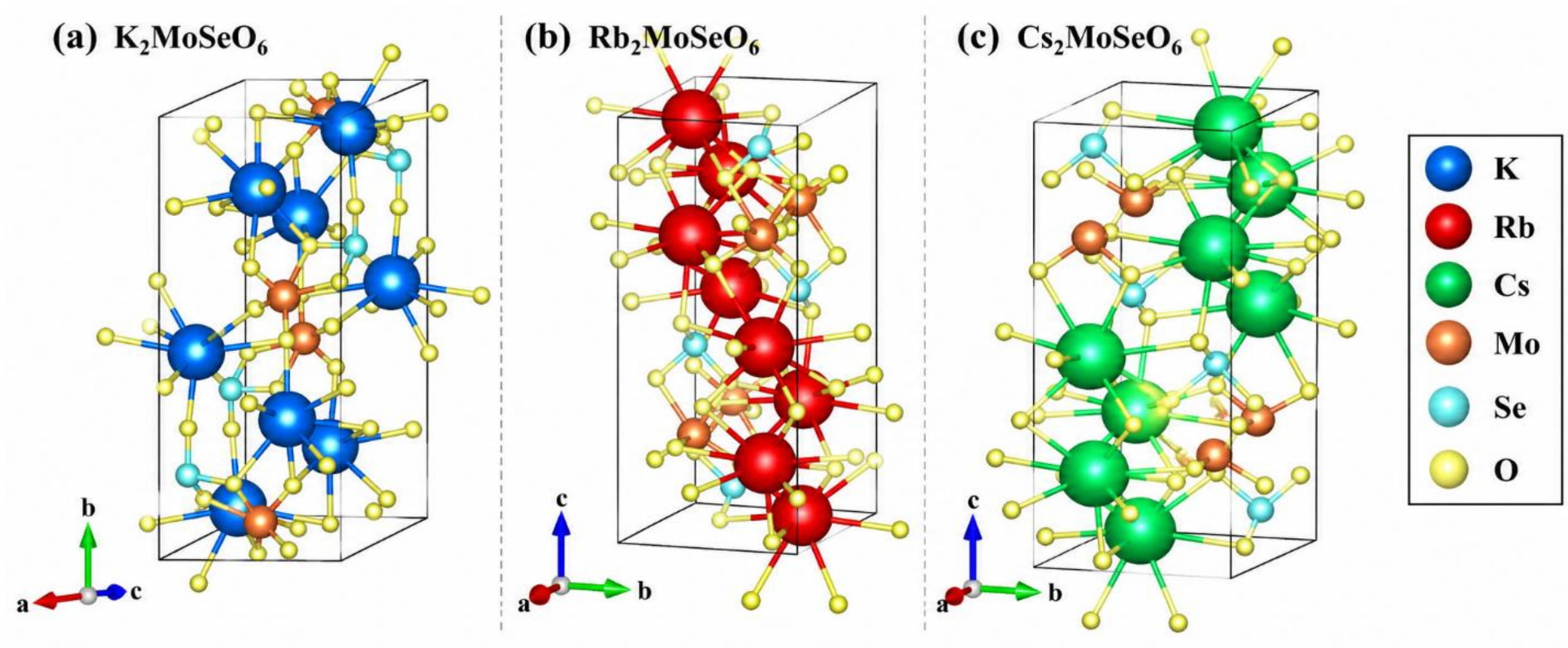

Fig. 1. Crystal structures of (a) $K_2MoSeO_6$ (monoclinic, $P2_1/c$), (b) $Rb_2MoSeO_6$ (orthorhombic, *Pnma*), and (c) $Cs_2MoSeO_6$ (orthorhombic, *Pnma*), showing the unit-cell boundaries and atomic arrangements. The colors represent K (blue), Rb (red), Cs (green), Mo (orange), Se (cyan), and O (yellow).

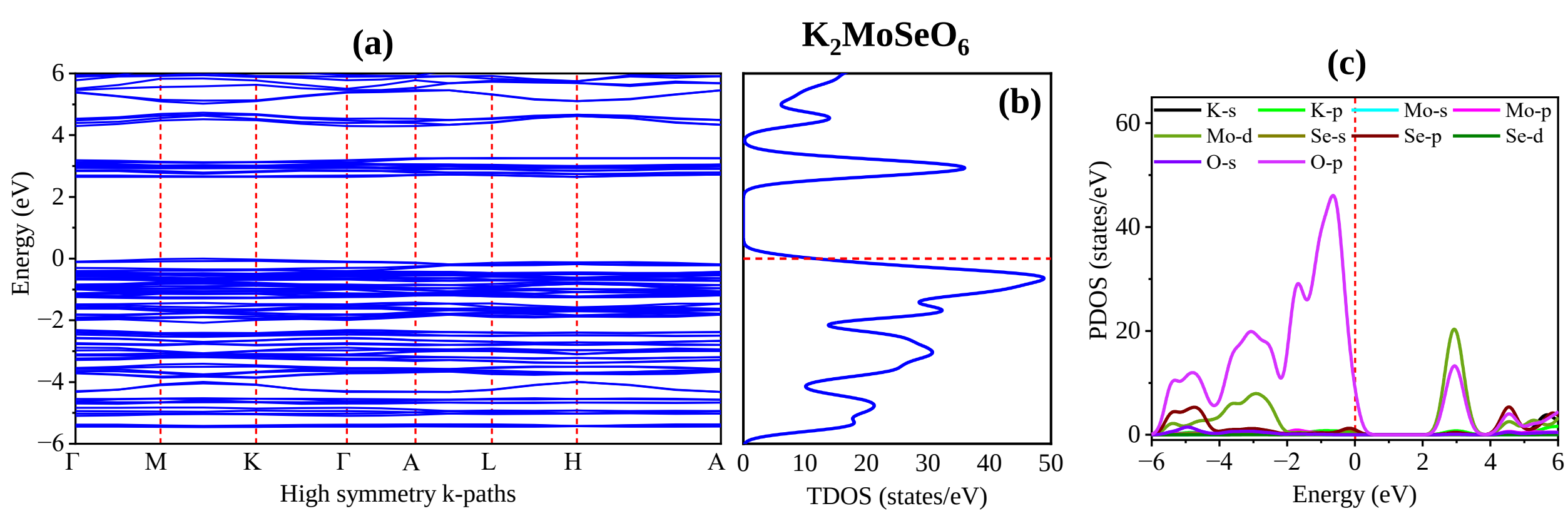

Fig. 2. Calculated (a) electronic band structures and (b) total density of states (TDOS) of $K_2MoSeO_6$, together with the corresponding orbital-projected density of states (PDOS) in (c). Red dashed lines indicate the zero-energy reference in the exported datasets.

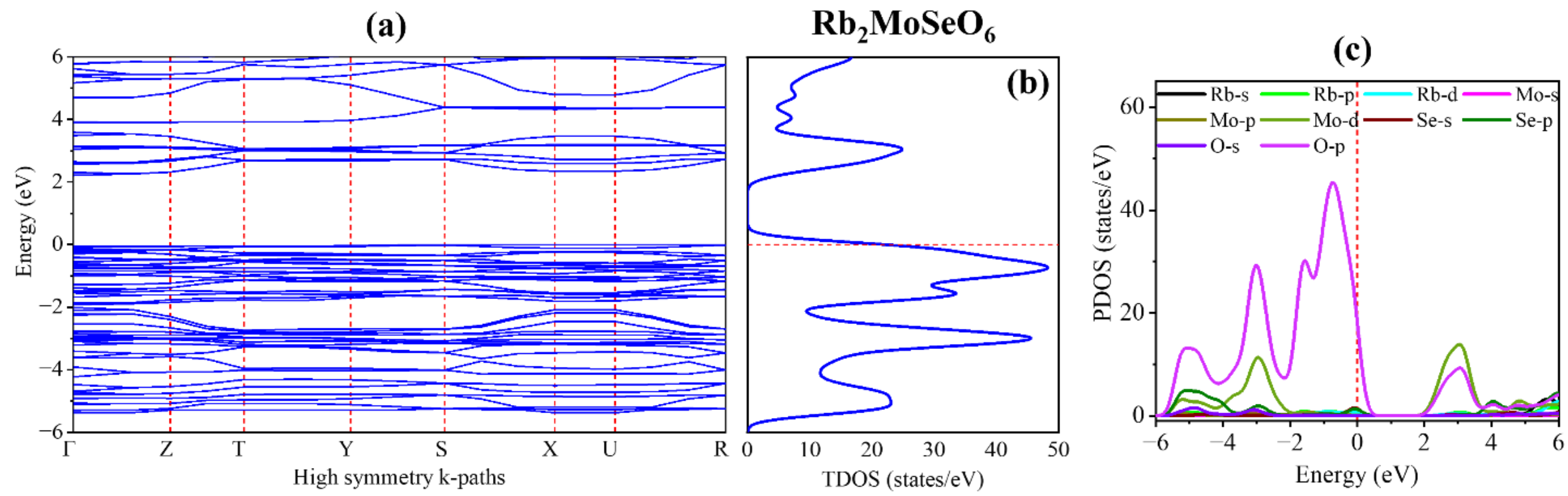

Fig. 3. Calculated (a) electronic band structures and (b) total density of states (TDOS) of $Rb_2MoSeO_6$, together with the corresponding orbital-projected density of states (PDOS) in (c). Red dashed lines indicate the zero-energy reference in the exported datasets.

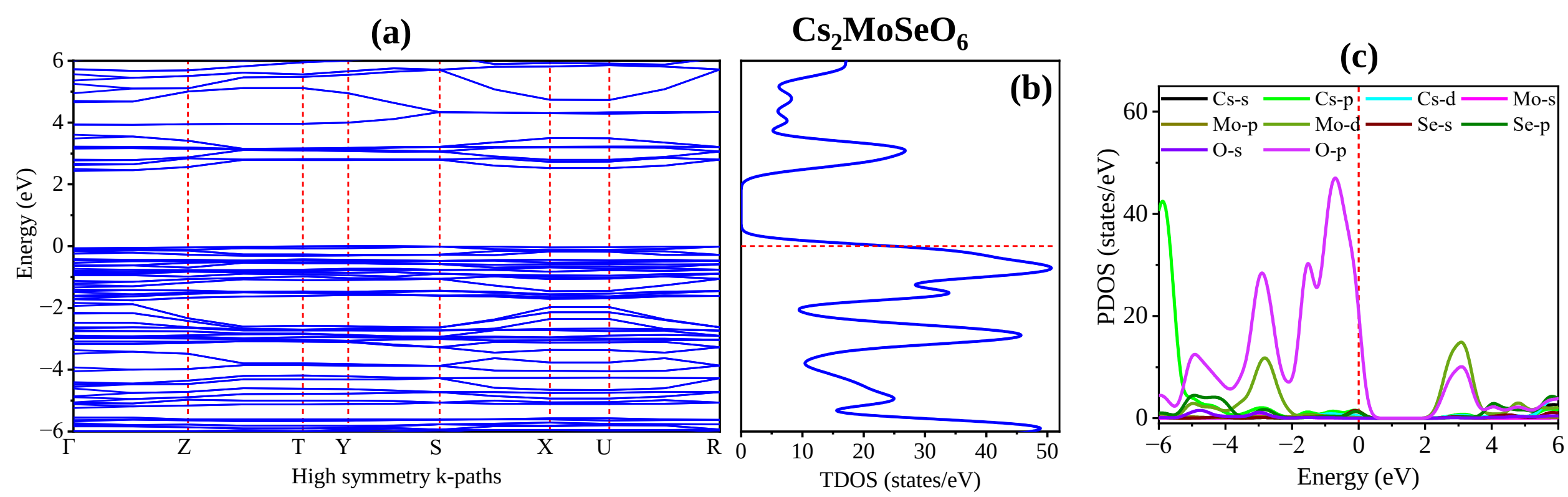

Fig. 4. Calculated (a) electronic band structures and (b) total density of states (TDOS) of $Cs_2MoSeO_6$, together with the corresponding orbital-projected density of states (PDOS) in (c). Red dashed lines indicate the zero-energy reference in the exported datasets.

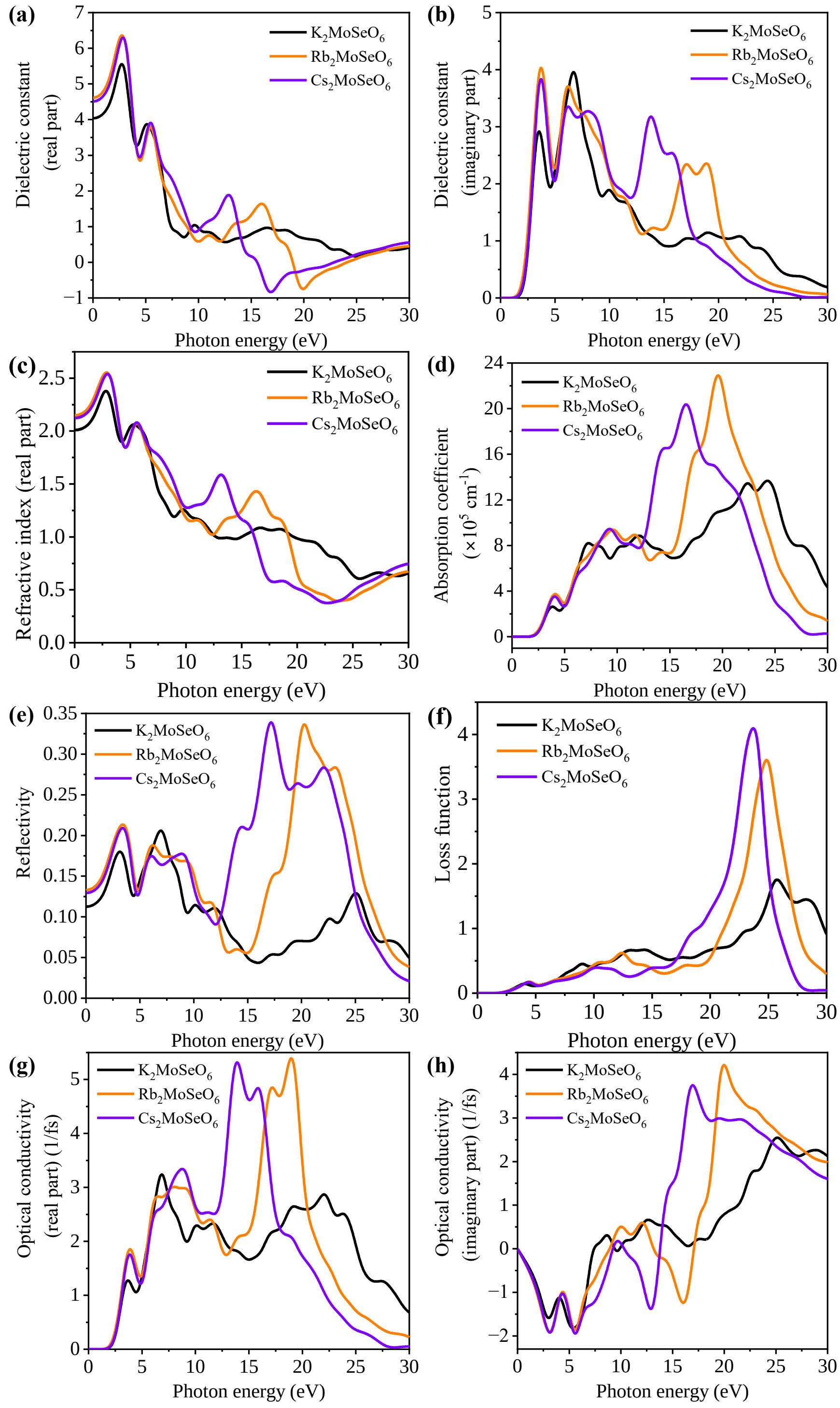

Fig. 5. Calculated photon-energy-dependent optical properties of X2MoSeO6 (X = K, Rb, and Cs): (a) real and (b) imaginary components of the complex dielectric function, (c) real part of the refractive index, (d) absorption coefficient, (e) reflectivity, (f) electron energy-loss function, and (g) real and (h) imaginary components of the optical conductivity. The black, orange, and violet curves correspond to K2MoSeO6, Rb2MoSeO6, and Cs2MoSeO6, respectively. The spectra illustrate the composition-dependent evolution of the dielectric response, interband optical transitions, absorption characteristics, and electronic energy-loss features over the photon-energy range of 0–30 eV.

## $K_2MoSeO_6$

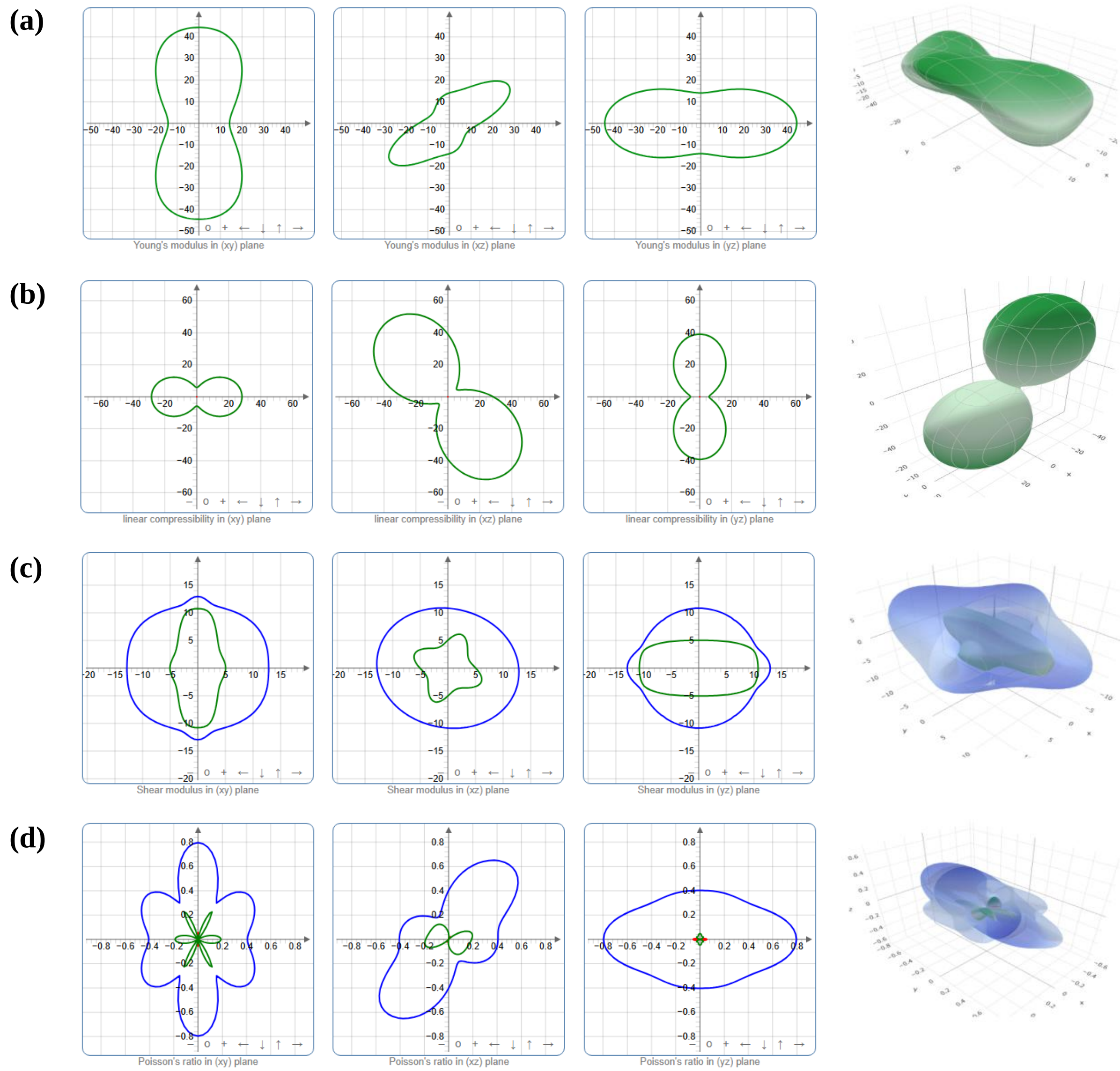

Fig. 6. Directional ELATE representations for $K_2MoSeO_6$: (a) Young's modulus, (b) linear compressibility, (c) shear modulus and (d) Poisson's ratio. Each row shows the supplied 2D polar plots and 3D surface.

## $Rb_2MoSeO_6$

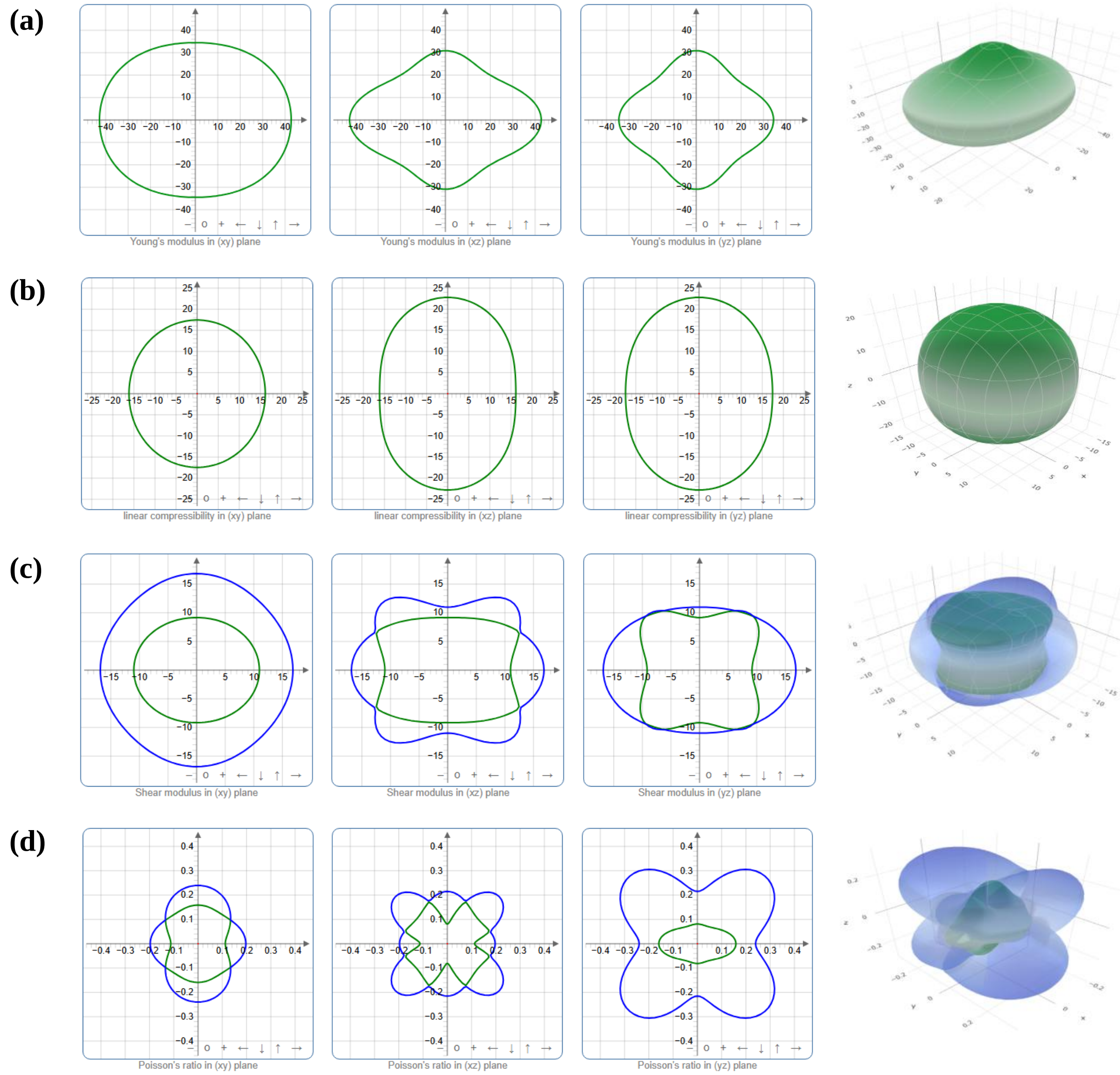

Fig. 7. Directional ELATE representations for $Rb_2MoSeO_6$: (a) Young's modulus, (b) linear compressibility, (c) shear modulus and (d) Poisson's ratio. Each row shows the supplied 2D polar plots and 3D surface.

**$Cs_2MoSeO_6$**

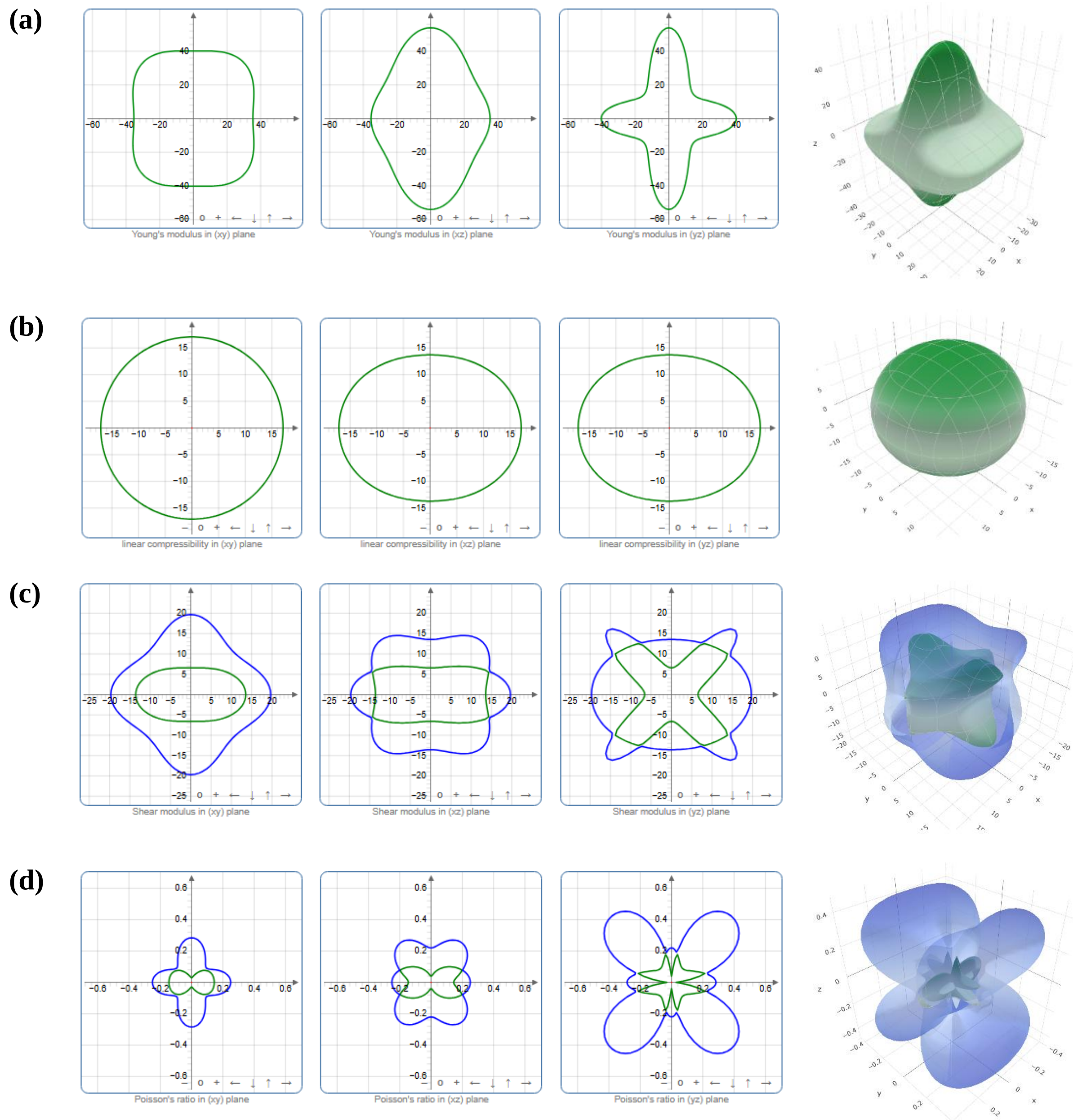

Fig. 8. Directional ELATE representations for $Cs_2MoSeO_6$: (a) Young's modulus, (b) linear compressibility, (c) shear modulus and (d) Poisson's ratio. Each row shows the supplied 2D polar plots and 3D surface.

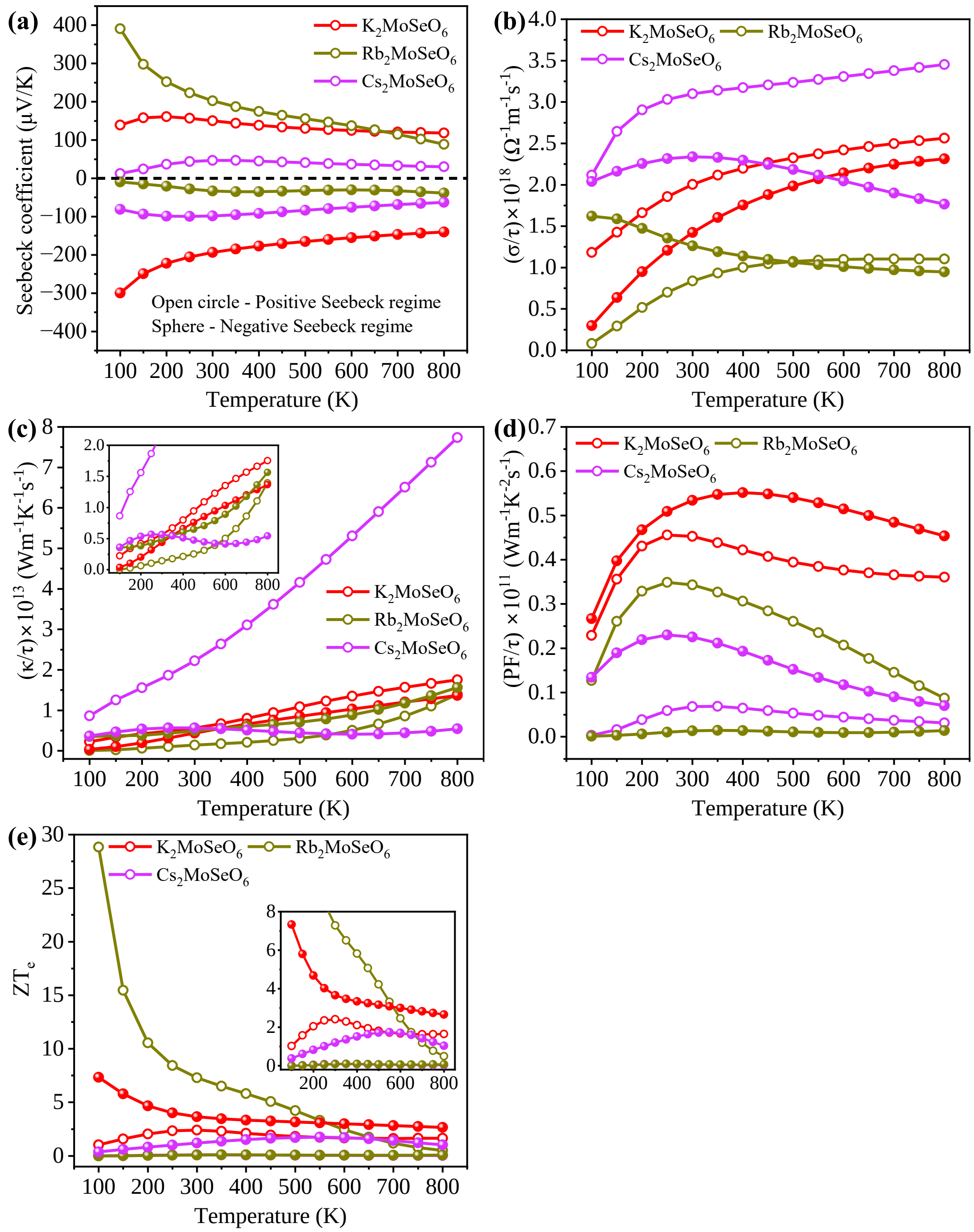

Fig. 9. Temperature-dependent thermoelectric properties of $K_2MoSeO_6$, $Rb_2MoSeO_6$, and $Cs_2MoSeO_6$: (a) Seebeck coefficient (S), (b) electrical conductivity per relaxation time (σ/τ), (c) electronic thermal conductivity per relaxation time ($\kappa_e/\tau$), (d) power factor per relaxation time (PF/τ), and (e) electronic-only figure of merit ($ZT_e$). For each compound, two chemical potentials corresponding to the maximum positive and most negative Seebeck coefficients at 300 K were selected to investigate the respective carrier-transport regimes. Open circles represent the positive-Seebeck regime (hole-dominated, p-type behavior), whereas spheres represent the negative Seebeck regime (electron-dominated, n-type behavior) across all panels. The insets in (c) and (e) provide enlarged views over reduced vertical-axis ranges to highlight variations that are less distinguishable in the main plots.